\documentclass[prb,aps,floatfix,twocolumn,superscriptaddress,footinbib,showpacs,citeautoscript]{revtex4-1}

\usepackage{graphicx}
\usepackage{epsfig}
\usepackage{braket}                                                                                                                                                                                                                                                                   
\usepackage{natbib}

\usepackage{amsfonts}
\usepackage{amsmath}
\usepackage{amssymb}
\usepackage{bm}

\usepackage[dvipsnames,usenames]{color}

\usepackage{comment} 
   
\begin{document}
\title{Magnetic ordering in quantum dots: Open vs. closed shells}
\author{J.~M. Pientka}
\affiliation{Department of Physics, St.~Bonaventure University, 
St.~Bonaventure, NY 14778}
\affiliation{Department of Physics, University at Buffalo, State University of New York, Buffalo, New York 14260, USA}
\author{R. Oszwa\l{}dowski}
\affiliation{Department of Physics, South Dakota School of Mines and
Technology, Rapid City, South Dakota 57701, USA}
\author{A.~G. Petukhov}
\thanks{Present address: Quantum Artificial Intelligence Laboratory, NASA Ames Research Center, 
Mail Stop 269-1, Moffett Field, CA 94035}
\affiliation{Department of Physics, South Dakota School of Mines and
Technology, Rapid City, South Dakota 57701, USA}
\author{J.~E. Han}
\affiliation{Department of Physics, University at Buffalo, State University of New York, Buffalo, New York 14260, USA}
\author{Igor \v{Z}uti\'{c}}
\email{zigor@buffalo.edu}
\affiliation{Department of Physics, University at Buffalo, State University of New York, Buffalo, New York 14260, USA}

\begin{abstract}
In magnetically-doped quantum dots changing the carrier occupancy, from open to closed shells, 
leads to qualitatively different forms of carrier-mediated magnetic ordering. While it is common 
to study such nanoscale magnets within a mean field approximation, excluding the spin fluctuations 
can mask important phenomena and lead to spurious thermodynamic phase transitions in small 
magnetic systems. By employing coarse-grained, variational, and Monte Carlo methods on singly 
and doubly occupied quantum dots to include spin fluctuations, we evaluate the relevance of the 
mean field description and distinguish different finite-size scaling in nanoscale magnets.
\end{abstract}
\maketitle 

\section{Introduction}
\label{sec:intro}

Nanoscale magnets are fascinating systems displaying phenomena at the boundary between 
classical and quantum physics. They reveal important implications for 
fundamental phenomena, 
such as macroscopic quantum tunneling\cite{Chudnovski1988:PRL,Friedman1996:PRL,Garanin1997:PRB},
magnetic polaron formation\cite{Seufert2001:PRL,Beaulac2009:S}, 
tunable magnetism\cite{Fernandez-Rossier2007:PRL, Abolfath2008:PRL,Abolfath2007:PRL}, and 
strongly-correlated states\cite{Oszwaldowski2012:PRB}, 
as well as 
potential 
applications in information storage and processing, 
arising from the superparamagnetic limit\cite{Wasner:2012}, magnetic hardening induced by 
nonmagnetic molecules\cite{Raman2013:N}, spin-lasers,\cite{Chen2014:NN,Zutic2014:NN}
and implementations of qubits.\cite{DiVincenzo 1995:S,Thiele2014:S} 

Despite the significant differences between nanoscale magnets and their bulk counterparts,
a mean-field description that could be suitable for bulk magnets in the thermodynamic limit,
remains also widely used in describing magnetic ordering in nanostructures. 
Unfortunately, the appealing simplicity of the mean-field approximation (MFA) 
can often mask important phenomena. Neglecting 
thermodynamic spin 
fluctuations can lead to spurious thermodynamic phase transitions in small magnetic systems.
Does that imply that the MFA cannot describe nanomagnets, or 
that there are situations
in which MFA could yield valuable and unexplored insights?
 
 \begin{figure}[h]
\centerline{\psfig{file=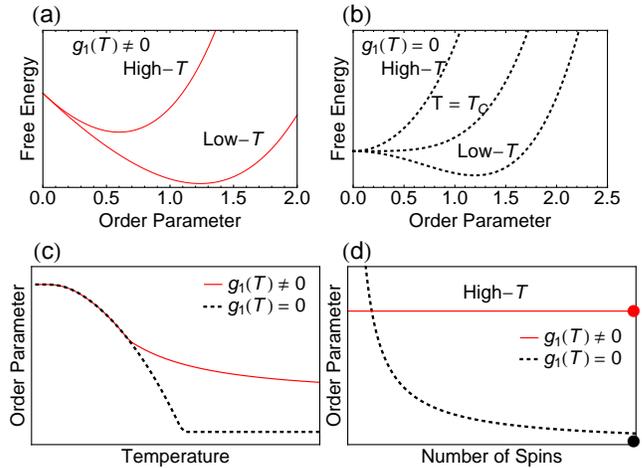,width=1\linewidth,angle=0}} 
\caption{(Color online)~Qualitative behavior of magnetic ordering in nanoscale systems. 
An example of a system that (a) does not 
and (b) does display a phase transition.  
The free energy, Eq.~(\ref{eq:freenergyfunc}), is shown as a function of 
the order parameter at various temperatures applicable to both (a) and (b).  
(c)  The temperature evolution 
of the  order parameter for the free energy 
described in (a) and (b).  
(d) For high temperature, there are two qualitatively different finite-size scalings 
for the normalized order parameter which also extrapolate to distinct thermodynamic limits (circles).} 
\label{fig:figINTRO1}
\end{figure}

In this work we show that the applicability 
of the MFA varies between different nanomagnets 
which also display different finite-size scaling and lead to distinct thermodynamic limits. 
We focus on magnetically-doped semiconductor quantum dots (QDs) with the localized 
impurity spins typically, provided by Mn ions.\cite{Maksimov2000:PRB,Seufert2001:PRL,Dorozhkin2003:PRB,
Holub2004:APL,Besombes2004:PRL,Hundt2005:PRB,Gurung2008:APL,Hoffman2000:SSC,Norris2001:NL,
Radovanovic2001:JACS,Beaulac2009:S,
Ochsenbein2009:NN,Bussian2009:NM,Zutic2009:NN,Viswanatha2011:PRL,Beaulac2011:PRB,Pandey2012:NN,
Farvid2014:JACS,Peng2014:JPC,
Xiu2010:ACSN,Sellers2010:PRB,Barman2015:PRB,Henneberger:2010,Klopotowski2011:PRB,Klopotowski2013:PRB,Pacuski2014:CGD,
Kudelsk2007:PRL,vanBree2008:PRB,Govorov2004:PRB}  
These systems are multi-carrier 
generalizations 
of the magnetic polaron 
formation\cite{Yakovlev:2010,Dietl1982:PRL,Wolff:1988,Durst2002:PRB,Furdyna1988:JAP,Nagaev:1983,Kasuya1968:RMP,
Dietl1995:PRL,Awschalom1986:PRL} 
that can be viewed as a cloud of localized impurity spins, 
aligned through exchange interaction with a confined carrier spin.
The  characteristic signatures of magnetic polarons
is the presence of high-temperature tails in the root mean square magnetization,    
rather than an 
abrupt vanishing of magnetization at the Curie temperature, $T_C$, 
expected for bulk magnets.\cite{Marder:2010}

To qualitatively distinguish magnetic ordering in different nanomagnets, 
such as epitaxially-grown or 
colloidal QDs,\cite{Hoffman2000:SSC,Norris2001:NL,Radovanovic2001:JACS,Beaulac2009:S,
Ochsenbein2009:NN,Zutic2009:NN,Bussian2009:NM,Viswanatha2011:PRL,Beaulac2011:PRB,Pandey2012:NN}
we introduce a simple description in which the relevant free
energy functional\cite{Dietl1982:PRL,Free_note} is reduced to a MFA free energy, 
$F$, given as a function of the order parameter 
$X$, and the absolute temperature, $T$,
\begin{equation}
F(X,T) = g_0(T) +  g_1(T) X +  \frac{ \alpha(T-T_C)}{2} X^2 +\frac{g_4(T)}{4} X^4.
\label{eq:freenergyfunc}
\end{equation}
where the expansion coefficients, $g_0$, $g_1$, $\alpha$,  and $g_4$ are functions of $T$. 
The quadratic and quartic terms in $X$ describe the entropy of 
the magnetic ions. 
Compared to the conventional Ginzburg-Landau form,\cite{Marder:2010} 
it is surprising to see in Eq.~(\ref{eq:freenergyfunc}) the linear term in $X$ in the absence
of an applied magnetic field, we describe the origin of this term in Sec.~\ref{sec:MP}. 

As shown in Fig.~\ref{fig:figINTRO1} (solid lines), the presence of a linear contribution in 
the order parameter, $g_1 \neq 0$, has a striking consequence. 
For {\em all} relevant  
$T$,\cite{temp_note} the free energy 
minimum is attained for a nonvanishing order parameter 
[Fig.~\ref{fig:figINTRO1}(a)], implying the absence of a spurious phase transition. 
The magnetic order remains finite for all 
$T$ [Fig.~\ref{fig:figINTRO1}(c)].
In contrast, for $g_1=0$ (broken lines) 
Figs.~\ref{fig:figINTRO1}(b) and (c) reveal a behavior typically associated with bulk ferromagnets
 displaying  a phase transition for $T>0$. 
 Surprisingly, this simple MFA description provides two qualitatively different finite-size scaling 
 with number of spins (magnetic ions) 
 in Fig.~\ref{fig:figINTRO1}(d)
 which we later show accurately reflects the behavior of  two classes of nanoscale magnets 
 by considering a more rigorous approach including spin fluctuations.

After this introduction, in Sec.~\ref{sec:theory} we provide an overview of the employed
theoretical methods and discuss the importance of the correct choice of the order parameter. 
We then focus on the two classes of nanomagnets and the 
simplest magnetic QD embodiment: (i) a single occupancy, 
implying the finite carrier spin configuration of an open shell in Sec.~\ref{sec:MP}, 
and (ii) a double occupancy,  corresponding to the vanishing carrier spin configuration of a closed shell
in Sec.~\ref{sec:MBP}.
We explain how these two classes of nanomagnets
are already qualitatively different at the MF level corresponding to (i) $g_1\neq0$ and (ii) $g_1=0$
behavior, 
in Fig.~\ref{fig:figINTRO1} and how they 
can be viewed as representing magnetic 
polarons\cite{Seufert2001:PRL, Beaulac2009:S,Maksimov2000:PRB,Sellers2010:PRB, Barman2015:PRB,Govorov2005:PRB,Govorov2008:CRP,Cheng2008:PRB} %I32 
and magnetic bipolarons,\cite{Oszwaldowski2012:PRB,Oszwaldowski2011:PRL,Bednarski2012:JPCM,Abolfath2012:PRL}
respectively. We conclude our presentation with the implications 
for the relevance of MFA to nanomagnets and discuss outstanding questions.

\section{Theoretical Overview}
\label{sec:theory}

\begin{figure}[h]
\centerline{\psfig{file=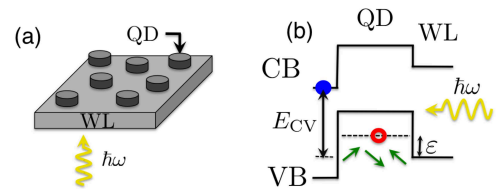,width=0.75\linewidth,angle=0}} 
\caption{(Color online)~(a) A scheme of QDs grown on a 
two-dimensional wetting layer (WL) in which 
electron-hole pairs are created by inter-band absorption of light.
Holes are subsequently captured in the QD.  
Type-II conduction/valence band (CB/VB) 
profile\cite{Kuskovsky2007:PRB} of a II-VI QD doped with Mn spins.\cite{Sellers2010:PRB} 
$\varepsilon$ and $E_{CV}$ are the confinement and bandgap energies. }
\label{fig:figINTRO2}
\end{figure}

Magnetically-doped QDs are 
a useful model system to study magnetic ordering 
in nanostructures.  
Even with very different growth techniques (top-town or bottom-up), such 
as epitaxially grown QDs or solution-processed colloidal QDs, there are striking similarities 
in the manifestations of their nanoscale magnetism, as well as in the limitations of their 
theoretical description.  We illustrate different implications of magnetic ordering by focusing 
on (II,Mn)VI QDs, depicted in Fig.~\ref{fig:figINTRO2}. These systems display carrier-mediated 
magnetism, extensively studied in bulk dilute magnetic semiconductors. Since Mn$^{2+}$ is isovalent 
with group II ions, carriers must be created independently; 
for example, excitation of electron-hole pairs by 
interband absorption of light [Fig.~\ref{fig:figINTRO2}(a)]. 
By changing the intensity of light 
can thus change the QD occupancy to realize both open- and closed-shell QDs.
A realization of multiple occupancy in QDs is 
observed in various 
experiments.\cite{Cherenko2010:PSSB,Gould2006:PRL,
Matsuda2007:APL,Bansal2009:PRB,Fisher2005:PRL,Besombes2005:PRB,Trojnar2013:PRB}

From the band alignment of conduction and valence bands in Fig.~\ref{fig:figINTRO2}(b), 
the ordering of Mn spins located in the QD region is dominated by the holes, characterized 
by the Mn-hole exchange coupling $\beta$. The electrons have a negligible influence on 
magnetic ordering. They are spatially removed from Mn and typically have $\sim$ 5-6 times 
smaller exchange coupling with Mn than holes.\cite{Furdyna1988:JAP}

\begin{figure}[h]
\centerline{\psfig{file=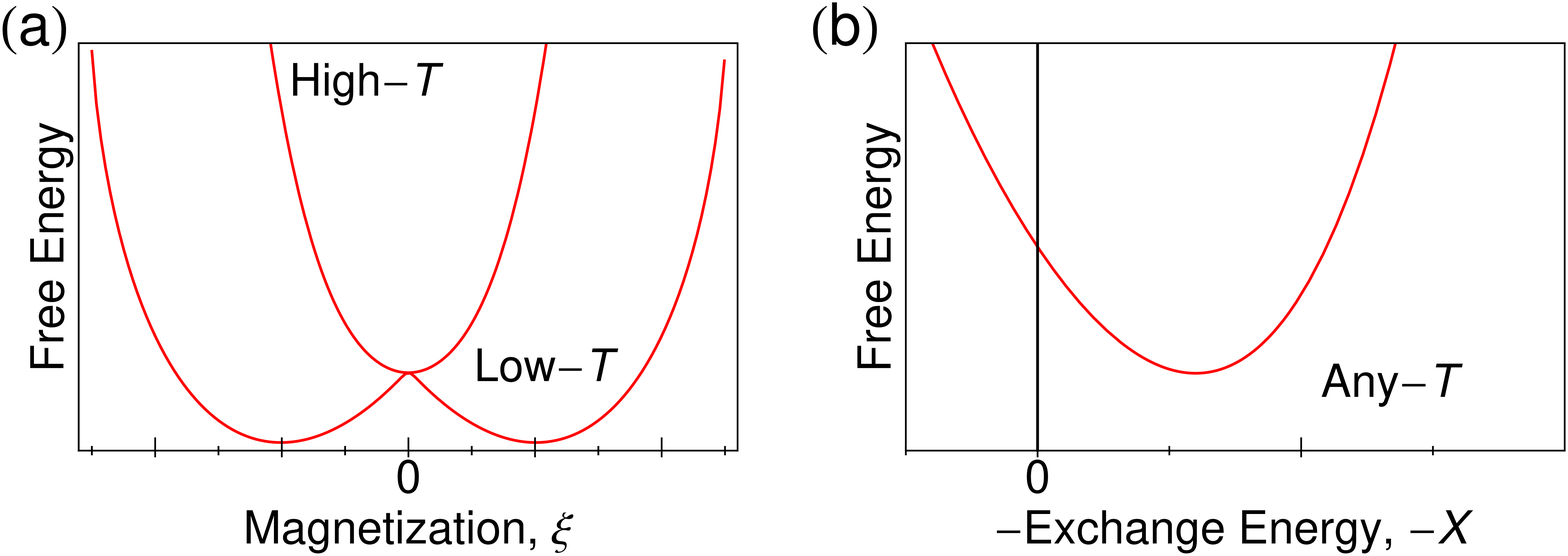,width=1\linewidth,angle=0}} 
 \caption{(Color online)~
 (a) The free energy 
 as a function of magnetization, $\xi$, 
 reveals a minimum at low-$T$  and vanishes 
 at high-$T$. 
 MFA 
 will give a second-order phase transition. 
(b) Corrected free energy 
of a nanoscale magnet as a function of  
``observable'' exchange energy, $X$, giving a finite value to the order for any finite $T$.  
 }
\label{fig:figfreeenergymethodmethods}
\end{figure}

We first recall the MFA 
as known from bulk systems, but due to its simplicity it is
often used in nanomagnets where its validity is questionable.  
A conventional mean-field theory  
is  
illustrated in Fig.~\ref{fig:figfreeenergymethodmethods},
where the most probable state of the system is obtained by minimizing 
the free energy as a function of the order parameter. 
For nanoscale systems, 
one needs to be 
careful with the choice of an order parameter. This 
often overlooked consideration, 
has striking implications, as depicted in Fig.~\ref{fig:figfreeenergymethodmethods}(a) and (b), 
 which we further explain on the examples of 
 magnetic polarons (MPs) and magnetic bipolarons (MBPs), discussed in 
 Secs.~\ref{sec:MP} and ~\ref{sec:MBP}.  A common choice of magnetization $\xi$ as the order 
 parameter, shown in Fig.~\ref{fig:figfreeenergymethodmethods}(a) leads to 
 spurious thermodynamic phase transition in nanoscale magnets.
In contrast, if an ``observable''
quantity, such as the 
exchange energy $X$, is chosen as the order parameter, the phase 
transition can be removed. 
This is shown in Fig.~\ref{fig:figfreeenergymethodmethods}(b), where there is always a 
minimum in the free energy 
at all $T$.  
Based on the choice of the order parameter we can then recover either 
the $g_1=0$ or the 
$g_1 \neq 0$
behavior of the free energy 
depicted in Figs.~\ref{fig:figINTRO1}(a) and (b).

Some qualitative trends in magnetic QDs can be obtained from the MFA developed for bulk dilute magnetic semiconductors.
The exchange coupling of a carrier (hole) and magnetic impurity (Mn) spins, $s$ and $S$, respectively, can be expressed 
in terms of two effective magnetic fields. The resulting self-consistent equations for the average spin 
(along the direction of the applied field or spontaneous Mn magnetization) are different depending 
on whether  the 
 carriers have nondegenerate or degenerate distribution~\cite{DasSarma2003:PRB,Petukhov2007:PRL}. 
 We will later discuss how these two bulk cases have 
 similarities with the  MFA  applied to MPs and MBPs.  

For nondegenerate carriers, the self-consistent equations are,\cite{Petukhov2007:PRL}
\begin{equation}
\langle s_z \rangle =s B_s \left( \frac{\beta a_0^3 n_i }{k_B T} \langle S_z \rangle  \right), 
 \label{eq:CarSpin}
\end{equation}
\begin{equation}
\langle S_z \rangle =S B_S \left( \frac{ \beta a_0^3 n_c}{k_B T} \langle s_z \rangle \right), 
 \label{eq:selfcon2}
 \end{equation} 
where $a_0^3$ is the unit cell volume, 
$n_{c}$ and $n_{i} $ 
are the carrier
and magnetic impurity 
densities, respectively; 
$k_B$ is the Boltzmann constant, and
 $B_J(x)$ is the Brillouin function, 
 \begin{equation}
B_J(x) = \frac{2J+1}{2 J}\coth \left( \frac{2J+1}{2J}x \right) - \frac{1}{2J}\coth \left( \frac{x}{2J} \right).
 \label{eq:Brill}
 \end{equation}
In the high-$T$ limit 
of a small ratio of the effective magnetic and thermal energies,
the expansion in Eq.~(\ref{eq:selfcon2}), 
 \begin{equation} 
 B_J(x) \approx \frac{J+1}{3J} x + O(x^3) \hspace{.2 in} x \ll1,
 \label{eq:BrillHighT}
 \end{equation}
 yields a vanishing magnetic response at a critical temperature, 
\begin{equation}
k_{\text{B}} T_C = \frac{1}{3 } \beta a_0^3 \sqrt{n_i n_c}\sqrt{S(S+1)s(s+1)}.
\label{eq:TCBULKnondeg}
\end{equation}

For the degenerate case, the carrier spin is given by 
\begin{eqnarray}
\langle s_z \rangle  &=& \frac{s}{2 n_c} \int d\varepsilon f(\varepsilon) \left[  D(\varepsilon + s \beta a_0^3 n_i \langle S_z\rangle ) \right. \nonumber \\
&-&\left.  D(\varepsilon - s \beta a_0^3 n_i \langle S_z\rangle ) \right].
\label{eq:CarSpindeg}
\end{eqnarray}
where $D(\varepsilon)$ is the density of hole states. 
In the high-$T$ limit, the integrand in Eq.~(\ref{eq:CarSpindeg}) can be expanded, using Eq.~(\ref{eq:selfcon2}), and Eq.~(\ref{eq:BrillHighT}), 
to give the critical temperature 
\begin{equation}
k_{\text{B}} T_C = \frac{1}{3}(\beta a_0^3)^2 n_i S(S+1)s^2  D(\mu),
 \label{eq:TCBULKdeg}
 \end{equation}
where $\mu$ is the chemical potential. 
Interestingly, the 
difference between the linear and quadratic $\beta$-dependence of the $T_C$  
in MFA  for the two 
bulk dilute magnetic semiconductors in Eq.~(\ref{eq:TCBULKnondeg}) and (\ref{eq:TCBULKdeg})
is also obtained using the MFA for MPs and MBPs, respectively.

 At low-$T$, where spin fluctuations are small, MFA can accurately 
 describe the thermodynamics of a finite-size system. However,
 a careful treatment is needed for  a higher $T$, 
 where large spin fluctuations could play  the  
 dominant role in the thermodynamics of  magnetic 
 QDs.\cite{Govorov2005:PRB,Pientka2012:PRB}

Unlike many studies of magnetic QDs that do not go beyond mean field theory,% 
\cite{Abolfath2007:PRL,Fernandez-Rossier2004:PRL,Lebedeva2010:PRB,Lebedeva2012:PSSB,Abolfath2012:PRL}
 we will utilize two methods which include spin fluctuations.
The first is a coarse-grained approach\cite{Oszwaldowski2012:PRB}  
in which we discretize the QD space into a number of cells, $N_c$
with $N_k$ being the number of Mn spins belonging to
each 
grid point
where 
$\sum_{j=1}^{N_k} S_{jz}$ is the projection of the total spin onto the 
$z$-axis of the Mn contained at the $k^{\text{th}}$ grid point. 
Within a given cell the wave function is slowly  varying allowing one to neglect the 
spatial dependence of the carrier and spin density.  
The full partition function is obtained by summing over all configurations of the 
normalized magnetization in a given cell. 

The second method 
is to perform Monte Carlo simulations. 
Unlike the coarse-grained method, Mn can be positioned at many sites allowing for 
spatial variation in the carrier spin density.  The Monte Carlo simulation seeks approximate 
solutions to the Schr\"{o}dinger equation, for the Hamiltonian $\hat{H}$, 
at a fixed $T$,  
for a 
finite orthonormal basis $\ket{\Phi}$ at a given Mn spin $\{S_z\}$ configuration. 
The calculation entails randomly generating a Mn configuration at a given $T$ and producing a matrix
representation of $\hat{H}(\{S_z\})$ in a finite basis and solving the eigenvalue problem.
A metropolis algorithm is used to obtain the most probable Mn configuration at a fixed $T$.\cite{Binder:2009}

Our model to study magnetic ordering in open- and closed-shell systems is motivated
by the Mn-doped QD with type-II band alignment in Fig.~\ref{fig:figINTRO2}, 
was shown experimentally to support
robust MPs.\cite{Sellers2010:PRB}
We use the total QD Hamiltonian,  $\hat{H}=\hat{H}_c+\hat{H}_{\mathrm{ex}}$, with typical two-dimensional (2D)
nonmagnetic (carrier) and magnetic (Mn-hole exchange) parts, where 
\begin{equation}
\hat{H}_c = \sum_{i=1}^{N_h}\left[-\frac{\hbar^2}{2m^*} \nabla_i^2+ \frac{1}{2}m^* \omega^2r_i^2 \right]+U_{N_h},
\label{eq:HKVs}
\end{equation}
$\hbar$ is the Planck's constant 
there are $N_h$ holes 
at the position ${\bf r}_i$ and $m^{\ast}$ is their 
effective mass.  A harmonic $x-y$  
confinement of strength $\omega$ is much weaker than the confinement 
along the growth ($z$) direction, implying effectively 2D system. $U_{N_h}$ is the charging energy. 
The $p-d$ exchange interaction between spins of Mn and confined holes 
has the Ising form\cite{Dorozhkin2003:PRB,Lee2014:PRB,Stano2013:PRB,Vyborny2012:PRB,Gall2011:PRL} 
because of the strong $z$-axis anisotropy, arising from spin-orbit interaction in the 2D QDs with energetically favorable heavy holes,
 \begin{equation}
\hat{H}_{\rm{ex}} = - \frac{\beta}{3} \sum_{i=1}^{N_h}\sum_{j=1}^{N_{\text{Mn}}} \hat{s}_{zi} \hat{S}_{zj}\delta({\bf r}_i - {\bf R}_j),
\label{eq:Hex}
\end{equation}
where there are $N_{\text{Mn}}$ Mn spins 
at the position ${\bf R}_j$.
Here,  $\hat{s}_{z}$ is the heavy-hole (pseudo)spin operator with projections $s_{z}=\pm 3/2$, while 
$\hat{S}_z$ is the operator of the $z$-projection of the Mn-spin $S=5/2$.  
Our theory does not include  antiferromagnetic interactions between neighboring Mn ions, 
which is relevant for QD's doped with large Mn concentrations\cite{Klosowsk1991:JAP}.    

Since $\hat{H}_{\mathrm{ex}}$ does not contain spin-flip processes, 
the total wave function is a product of the hole and Mn-spin parts.\cite{Wolff:1988}
The partition function of the system can be calculated using a Gibbs canonical distribution, 
\begin{equation}
Z=\text{Tr}_{S_{jz},s_{zi}}e^{-\hat{H}/k_{\text{B}} T },
 \label{eq:gibss}
 \end{equation}
which even for a single hole has a prohibitive complexity to be solved exactly.  
To calculate $Z$, in a typical QD with $N_{\text{Mn}} \sim 100-1000$  and $S$=5/2 for Mn spin,
one would need to solve $6^{N_{\text{Mn}}}$  replicas of the hole Schr\"odinger equation.

To overcome this computational complexity and gain insight in magnetic ordering of
open and closed-shell QDs we first use the MFA.  We next consider
a coarse-grained approach of discretizing the QD space to include spin fluctuations and 
examine the limitations of the MFA. We investigate the thermodynamics of the MP and the 
MBP formation and explore the finite-size effects by varying the number of magnetic 
impurity spins. We further corroborate our results and the influence of spin fluctuations
using Monte Carlo simulations.

\section{Magnetic Polarons (MP\MakeLowercase{s})}
\label{sec:MP}

Early studies of MPs considered bulk magnetic semiconductors where the localized carrier spin
was provided by the donor or acceptor.\cite{Nagaev:1983,Kasuya1968:RMP} 
For such bound magnetic polarons a finite 
extent of the carrier wave function  leads to the alignment of only a small number of a nearby spins of magnetic
impurities having many similarities with MPs in magnetic QDs. While problems of 
a conventional MFA in describing experiments on bound magnetic polarons have been 
explained 
over thirty years ago,\cite{Dietl1982:PRL}  (spurious critical 
behavior and thermodynamic phase transitions in very small magnetic systems was removed
after including spin fluctuations) such pitfalls  continued to be repeated in describing magnetic QDs.

We begin by considering a singly occupied QD [$N_h = 1$, $U_{N_h}=0$, recall Eqs.~(\ref{eq:HKVs}) and (\ref{eq:Hex})], the 
simplest realization of an open-shell QD,\cite{note_open} 
 and study the thermodynamics of the MP.  
We build the partition function by constructing a canonical Gibbs distribution, 
Eq.~(\ref{eq:gibss}).
The distribution
function, $\Omega_S(N_k, \xi_{k})$, that describes the number of 
configurations of free spins in a given cell expressed in terms of the 
microscopic parameter, 
$\xi_{k}$ (which can be viewed as a normalized magnetization),
\begin{equation}
\Omega_S(N_k,\xi_k) = \sum_{\left\{  S_{jz}\right\}  }\ \delta \left( \xi_k - \frac{1}{N_k S} \sum_{j=1}^{N_k} S_{jz}  \right), 
\label{eq:OMEGA}
\end{equation} 
where (recall Sec.~\ref{sec:theory})  $\sum_{j=1}^{N_k} S_{jz}$ is the projection of 
the total Mn 
spin onto the $z$-axis,
and the argument of the $\delta$-function defines the normalization of $\xi$  
while the $\delta$-function is the distribution of Mn spins in a cell.  We find $\Omega_S(N_k, \xi_k)  
 \propto \exp[- G_S(\xi_k)/k_{\text{B}}T]$ 
(see Appendix~\ref{MnDist} for details) with
\begin{equation}
G_S(\xi_k)=k_{B} T N_k  \left [\xi_k B_S^{-1}\left(\xi_k \right) - \ln Z_S\left(B_S^{-1}\left(\xi_k \right)\right)\right] ,
\label{eq:MnFree}
\end{equation}
being the free energy 
for non-interacting spins, 
where $Z_S(x) = \sinh\left[ (1+1/2S) x \right] /\sinh\left[ x /2S \right]$, $B_S^{-1}(y)$ 
is the inverse of the Brillouin function $y=B_S(x)$. \cite{Kubo:1960}
 \begin{figure}[h]
\centerline{\psfig{file=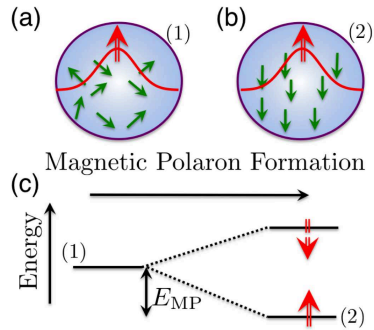,width=0.6\linewidth,angle=0}} 
\caption{(Color online)~ (a) A magnetic QD doped with random paramagnetic Mn ions (green arrow) 
with one carrier spin density (red arrow).
(b) The system lowers its energy through the exchange 
interaction and results in an 
antiferromagnetic 
alignment between the hole spin and the
Mn spins producing the MP. (c) The doubly degenerate QD 
energy level (1) splits with formation of the MP (2).  
The difference between the nonmagnetic QD energy 
and the ground state energy is the average exchange energy, $E_{\text{MP}}$. } 
\label{fig:MPFORM.png}
\end{figure} 

A simple manifestation of a magnetic ordering in an open-shell QD occupied by a single carrier 
is the MP formation depicted in Fig.~\ref{fig:MPFORM.png}.
Through exchange interactions between the carrier and Mn spins, once random paramagnetic
Mn ions [Fig.~\ref{fig:MPFORM.png}(a)] acquire their spin alignment [Fig.~\ref{fig:MPFORM.png}(b)]
(antiferromagetically coupled  to a hole spin) and reduce the total energy of the carrier-Mn system. 
As a result of the hole-Mn exchange term in Eq.~(\ref{eq:Hex}), the doubly degenerate heavy-hole 
energy level is split into two 
nondegenerate energy levels, as shown in Fig.~\ref{fig:MPFORM.png}(c). 
This corresponds to a red shift of the interband transition 
energy as a function of time, which is observed in time-resolved photoluminescence 
experiments.\cite{Seufert2001:PRL,Beaulac2009:S,Sellers2010:PRB,Barman2015:PRB}

In discrete space, for a given configuration of Mn  spins, $\{\xi \}$,
the two energy eigenvalues of Eq.~(\ref{eq:Hex}) are, 
$E_{\pm}= \pm \Delta_{\text{MP}}/2$, 
with spin-splitting energy
\begin{equation}
\Delta_{\text{MP}} = \frac{2\beta}{3}S \sum_k N_k   \rho_{\text{MP}}(R_k) \xi_k, 
\label{eq:gammadef}
\end{equation}
where,
\begin{equation}
 \rho_{\text{MP}}(R_k) =3|\phi (R_k)|^2/2,   
\label{eq:rhoMP}
\end{equation}
is the heavy hole spin density at the $k$'th cell, expressed in terms of the corresponding
wave function $\phi$. 

A 
MFA  result for the spin-splitting energy $\Delta^{\rm MF}_{\text{MP}}$, is  
obtained 
by inter-relating the carrier and Mn spin densities $\langle s_z \rangle$ and $\langle \xi_k \rangle$,
respectively,
in analogy of Eqs.~(\ref{eq:CarSpin})
and (\ref{eq:selfcon2}),
\begin{equation}
\langle{s_z}\rangle =\frac{3}{2}\tanh \left( \frac{3}{2} \sum_k  \frac{\beta |\phi(R_k)|^2}{3 k_B T}N_k S \langle \xi_k \rangle     \right),
\label{eq:szMFMP}
\end{equation}
and
\begin{equation}
\langle \xi_k \rangle = B_S\left( S \frac{\beta |\phi(R_k)|^2  }{3 k_B T} \langle s_z \rangle \right).
\label{eq:SMFMP}
\end{equation}
Substituting Eq.~(\ref{eq:szMFMP}) and (\ref{eq:SMFMP}) into Eq.~(\ref{eq:gammadef}) gives

 \begin{equation}
\frac{\Delta^{\rm MF}_{\text{MP}}}{2}=\frac{\beta}{3}S\sum_k  N_k \rho_{\text{MP}}(R_k)
B_{S} \left( \frac{ \beta S \rho_{\text{MP}}(R_k)}{3 k_B T} \tanh \left[ \frac{\Delta^{\rm MF}_{\text{MP}}}{2 k_B T} \right]\right). 
 \label{eq:MPMFSELFCON}% 
\end{equation}
In the unsaturated limit, $\Delta^{\rm MF}_{\text{MP}}/2 k_B T \ll1$, Eq.~(\ref{eq:MPMFSELFCON}) 
 gives a vanishing $\Delta^{\rm MF}_{\text{MP}}$ at a critical temperature,
 \begin{equation}
 k_{\text{B}} T_{C,{\rm MP}}^{\rm MF}=\frac{\beta}{3} \left(\frac{S(S+1)}{3}\sum_k N_k \rho^2_{\text{MP}}(R_k) \right)^{1/2}.
 \label{eq:MPMFSELFCONTC}% 
\end{equation}
The MP energy, $E_{\text{MP}}$, is defined as the expectation value of Eq.~(\ref{eq:Hex}). 
The MFA 
expression  
is 
 \begin{equation}
 E_{\text{MP}}^{\text{MF}}= -\frac{\Delta^{\rm MF}_{\text{MP}}}{2}\tanh \left( \frac{\Delta^{\rm MF}_{\text{MP}}}{2 k_B T} \right).
  \label{eq:EMPMF} 
\end{equation}

Having derived a MF theory
description for the MP, one can make a connection between the MP and
to that of a 
bulk nondegenerate magnetic semiconductors. 
Comparing Eqs.~(\ref{eq:MPMFSELFCONTC}) and (\ref{eq:TCBULKnondeg}), we see 
in both cases that the critical temperature
has a linear dependence on $\beta$.
In a nondegenerate bulk semiconductor the concentration of donor/acceptor atoms is so small that the Pauli exclusion principle
is ineffective and does not 
alter the carrier spin distribution which is described by the Boltzmann statistics. 
The magnetic impurity spins tend to align with one carrier spin and form a bound magnetic polaron.\cite{Nagaev:1983} 
The carrier spin in a nondegenerate magnetic semiconductor does not interact with other carriers spins.  
Thus, like the carrier in a magnetic QD, the carrier spins are free spins and will tend to align the spins of magnetic impurities.      

 \begin{figure}[h]
\centerline{\psfig{file=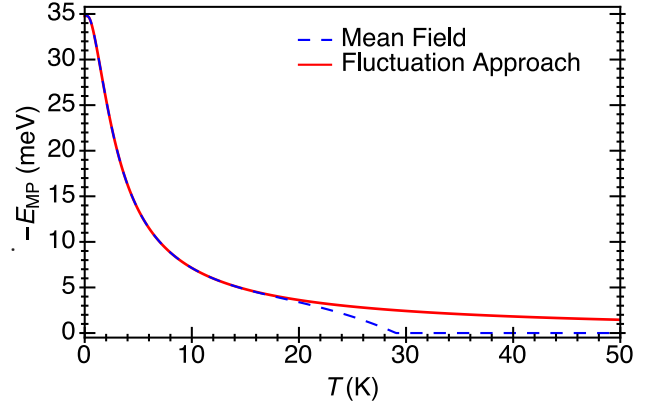,width=1.0\linewidth,angle=0}} 
\caption{(Color online)~ A mean field solution for MP blue/dashed curve and the result obtained 
through the fluctuation approach red/solid. For simplicity, we approximate the hole 
wave function as uniform throughout the QD volume $V$, $\phi(r) = 1/\sqrt{V}$. 
\cite{Golnik1983:JPC} The parameters are $N_0\beta$ = $-$1.05 eV,  
the cation density, $N_0=4/a_0^3$ with $a_0 \simeq 6.1~$\AA, $V = \pi r^2 h_z$, $r=5$~nm, $h_z=2.5$~nm, 
 $N_{\text{Mn}}= 90$.   }
\label{fig:figMFFAMP}
\end{figure}

Next we turn our analysis to the MP energy.
Motivated by typical QD parameters,\cite{Pientka2012:PRB} the blue/dashed line in Fig.~\ref{fig:figMFFAMP} 
shows the MF behavior of the MP energy Eq.~(\ref{eq:EMPMF}) as a function of $T$.  
For the chosen parameters, MF predicts a second order phase transition at characteristic temperature $T_{C,{\rm MP}}^{\rm MF}=29$~K.  
The phase transition occurs at a 
$k_B T$  
comparable to the exchange splitting of hole
levels.  At this temperature 
both hole states of opposite spin are approaching each other with equal probability.  
Consequently, yielding vanishing average MF spin density and average exchange energy  
at a finite $T$. 

MFA neglects the possibility for the system to deviate
from the minimum configuration, $\{\xi_k^{\rm MF}\}$,
leading to phase transitions not allowed in nanoscale systems.
To demonstrate the removal of the phase transition, we use the full partition function 
for the MP by summing over all configurations of $\{\xi_k\}.$\cite{Dietl1982:PRL}
The partition function is
\begin{equation}
Z_{\text{MP}}=\sum_{\sigma=\pm 1}\int
\exp\left(\frac{\sigma\Delta_{\text{MP}}[\xi]}{2k_B T}\right) 
\prod_{k=1}^{N_c} \Omega_S(N_k, \xi_{k}) d^{N_c}\xi,
\label{eq:PartMP1}
\end{equation}
with the spin index $\sigma=\pm 1$ for the heavy-hole spin
$s_z=\frac32\sigma$. Since the distribution $\Omega_S(N_k, \xi_{k})$ is
an even function of $\xi_{k}$, 
the integrals for $\sigma=\pm 1$ yield
the same result and  
\begin{equation}
Z_{\text{MP}}=2\int
\exp\left(\frac{\Delta_{\text{MP}}[\xi]}{2k_B T}\right)  \prod_{k=1}^{N_c}  
\Omega_S(N_k, \xi_{k}) d^{N_c}\xi.
\end{equation}
Here the summation of $\sigma=\pm 1$ is done exactly, without neglecting
the statistical correlation between $\sigma$ and $\{\xi_k\}$ as in the MFA. 
Now, using the steepest descent method 
as above, we
have
\begin{equation}
Z_{\text{MP}}=2 \prod_{k=1}^{N_c} Z_S\left(   \frac{ \beta S
\rho_{\text{MP}}(R_k)}{3 k_B T} \right)^{N_k}.
\label{eq:PartMP2}
\end{equation}
The average exchange energy can be evaluated from $E_{\rm MP}=-k_BT\beta
d(\ln Z_{\rm MP})/d\beta$ as
\begin{equation}
E_{\text{MP}}= - \frac{\beta}{3}S\sum_k  N_k   \rho_{\text{MP}}(R_k)  B_S \left( \frac{  \beta S \rho_{\text{MP}}(R_k)}{3 k_B T}  \right),
 \label{eq:EMPp4}
\end{equation}
where the corresponding results from  Eq.~(\ref{eq:EMPp4}) include the fluctuations of Mn  spin and are compared 
in Fig.~\ref{fig:figMFFAMP} (red/solid) with the MFA results (blue/dashed). 
In many colloidal QDs the number of magnetic impurities is much smaller than used in Fig.~\ref{fig:figMFFAMP}
($N_\text{Nm}=90$), enhancing the importance of the fluctuations and the corresponding difference 
from the MFA solution.

\begin{figure}[h]
\centerline{\psfig{file=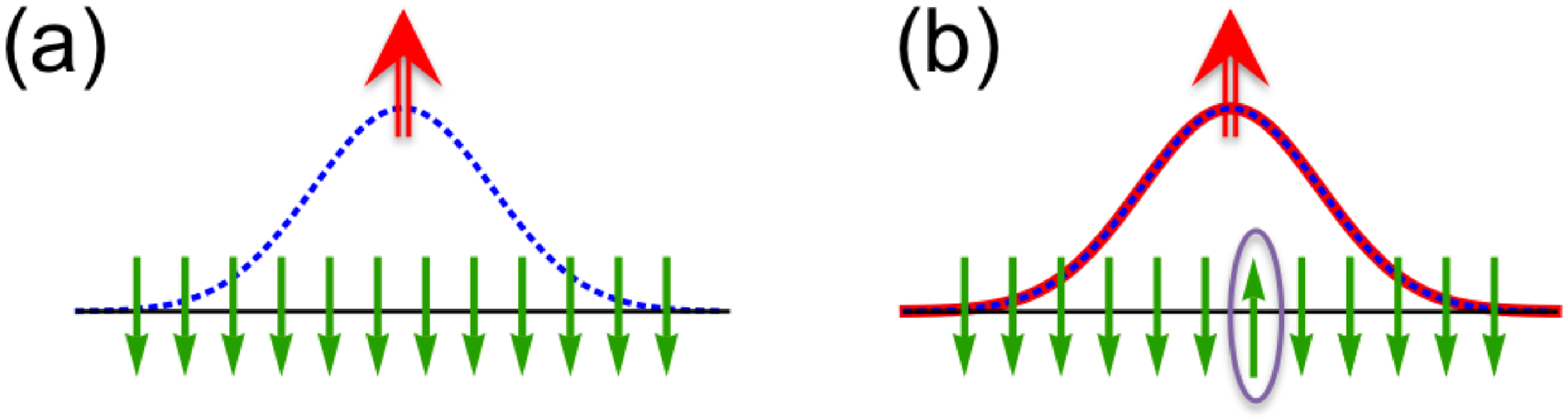,width=1.0\linewidth,angle=0}} 
\caption{(Color online)~Illustration of the finite-size effect for the MP 
(a) Spin density (blue/dashed) for fully aligned Mn spins (green). 
(b) A flipped Mn spin 
does not effect the carrier spin density (red/solid). 
 }
\label{fig:figMPFINITEEF}
\end{figure}

The MP shows different behaviors in the high- and low-$T$  
limits
\begin{equation}
E_{\rm MP}\propto\left\{
\begin{array}{ll}
-\beta & \mbox{ for }T\to 0 \\
\\
-\beta^2/k_BT & \mbox{ for }\beta S\rho_{\rm MP}(R_k)/k_BT\ll1, 
\end{array}
\right.
\end{equation}
which correspond to saturated and
unsaturated limits of magnetization, respectively. 
As shown in Fig.~\ref{fig:figMFFAMP}, $E_{\rm MP}$ has the $1/T$
behavior for a wide range of temperatures.
This can be understood as
follows. As depicted in Fig.~\ref{fig:figMPFINITEEF}, a single carrier
with uncompensated spin couples to 
a sum of many Mn spins.  
Therefore the
carrier aligns with the majority of Mn spin, 
and a flip of an
individual Mn spin would not affect the carrier spin, and consequently
other Mn spins. 
For this reason, the Mn spins 
are in effect
weakly interacting 
which results in a 
$1/T$ Curie-like temperature dependence of $E_{\rm MP}$. 

\begin{figure}[h]
\centerline{\psfig{file=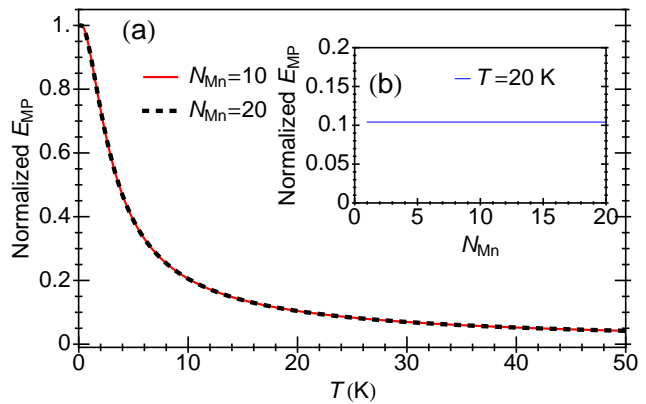,width=1.0\linewidth,angle=0}} 
\caption{(Color online)~ (a) The MP average exchange energy normalized by its $T=0$~K value for 
10 Mn (red/solid) and 20 Mn (black/dashed) as function of $T$. 
(b)  Normalized MP average exchange energy as a function of the total number of Mn at fixed $T$.}
\label{fig:figEMPNMN}
\end{figure}

We now show that 
the MP does not display a finite-size effect, when in a fixed QD volume we 
change the number of Mn spins. 
Using the full MP partition function and the resulting Eq.~(\ref{eq:PartMP2}), in 
Fig.~\ref{fig:figEMPNMN}(a) we show $E_{\text{MP}}(T)$, 
normalized with respect to its fully saturated value 
for various number of Mn.  Figure~\ref{fig:figEMPNMN}(b)
shows the normalized $E_{\text{MP}}$ plotted at fixed $T$ for various number of Mn ions.  
The normalized  $E_{\text{MP}}$ remains constant. The finite size-effect was 
accurately predicted by the MFA.  At the saturated and unsaturated limit, 
$E_{\rm MP}^{\rm MF} \propto N_{\text{Mn}}$. 
In Appendix~\ref{RENORM} we show how $E_{\rm MP}$ will depend on a different choice 
of a carrier wave function.

To summarize, we  
interpret the previous calculations in terms of the Ginzburg-Landau
approach of phase transitions. 
While the approach is not reliable due to the nanoscale
size of our system, we may gain important insights. 
Typically the partition function is described by a functional
integral over the magnetization $\xi$ 
[see Eq.~(\ref{eq:PartMP1})].
In the low-$T$ limit 
(or $|\Delta_{\text{MP}}|/2k_B T \gg1$)
with $e^{\Delta_{\text{MP}}/2k_B     
T}+e^{-\Delta_{\text{MP}}/2k_B T} \approx e^{|\Delta_{\text{MP}}|/2k_B
T}$ in Eq.~(\ref{eq:PartMP1}), 
one can write  
the free energy 
\begin{equation}
F_{\text{MP}}(\xi) \approx -g_1 |\xi| - k_B T S(\xi), 
\end{equation}
with the entropy $S(\xi)$ even in $\xi$. $F_{\text{MP}}(\xi)$ is depicted in Fig.~\ref{fig:figfreeenergymethodmethods}(a).
However, the finite potential barrier separating the degenerate minima at  $\pm \xi_{\text{min}}$ does
not prevent the thermal fluctuations between the local minima. Therefore the
correct solution $\langle \xi \rangle=0$ is not predicted by the 
MFA 
to  $F_{\text{MP}}(\xi)$.

We consider 
 the Ginzburg-Landau approach defined with a different
variable, the observable quantity of the exchange energy 
$X = E_{\text{ex}}=\sigma \Delta_{\text{MP}}/2$ 
[see Eq.~(\ref{eq:gammadef})] with the carrier spin index $\sigma = \pm 1$. 
The linear dependence in $\xi$ originates
from the finite carrier spin of the open shell.
Since the
carrier spin does not contribute to the entropy, the entropy in $X$ is directly
related to $S(\xi)$, and we derive the free energy 
in the order parameter $X = E_{\text{ex}}$ as
\begin{equation}
F_{\text{MP}}(X) \approx X- T S(X), 
\label{eq:FMPX}
\end{equation}
as depicted in Fig.~\ref{fig:figfreeenergymethodmethods}(b). 
Again, the entropy $S(X)$ is an even function of $X$ and
Eq.~(\ref{eq:freenergyfunc}) results for the magnetic polaron. Unlike $F_{\text{MP}}(\xi)$, $F_{\text{MP}}(X)$ 
possesses only one global minimum at a negative finite $X$ and $\langle X \rangle<0$ at all $T$, 
and the mean-field interpretation of Eq.~(\ref{eq:FMPX}) gives a qualitatively
correct prediction. The thermodynamic solution of a 
finite $\langle X \rangle $ manifests itself 
in the finite-scale independence in $\langle X \rangle /N_{\text{Mn}}$ in Fig.~\ref{fig:figINTRO1}(d).

\section{Magnetic Bipolarons (MBP\MakeLowercase{s})} 
\label{sec:MBP}

We next  
turn to 
a magnetic QD containing two holes [$N_h=2$, $U_{2}=U$,  
Eqs.~(\ref{eq:HKVs}) and (\ref{eq:Hex})]. 
Closed-shell fermionic systems, such as noble gases, are known for their stability and the
total spin-zero ground state, making them magnetically inert.
Thus it would seem that this simple example of a two-hole closed-shell QD doped 
with Mn would not allow magnetic ordering. However, the Mn doping does alter the magnetic 
properties of closed-shell QDs. The corresponding ground state, 
which is neither a singlet nor a triplet,
allows ordering of Mn spins, owing to the spontaneously broken time-reversal symmetry.\cite{Oszwaldowski2011:PRL}
 
\begin{figure}[h]
\centerline{\psfig{file=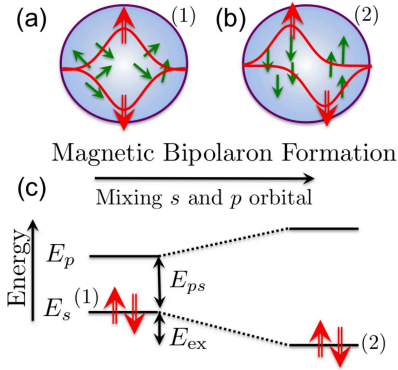,width=0.6\linewidth,angle=0}} 
\caption{(Color online)~  (a) A magnetic QD doped with random 
paramagnetic Mn ions spins (green arrows).
(b) The system lowers its energy through the exchange interaction leading to a 
nonvanishing 
spin density and formation of the MBP. 
Red arrows and lines in (a) and (b) show how the two-carrier spin density changes due to the presence Mn. 
(c) The doubly occupied QD energy level (1) lowers its energy through the formation of  the MBP (2).  
The difference between the nonmagnetic QD energy and the MBP ground state energy is the 
average exchange energy $E_{\text{ex}}$.}
\label{fig:MBPFORM}
\end{figure}

To lower the hole-Mn system energy through exchange interaction, there needs to be a nonvanishing
hole spin density.  During the MBP formation, an initially random Mn spin orientation [Fig.~\ref{fig:MBPFORM}(a)], 
in the presence of two holes acquires a spin alignment [Fig.~\ref{fig:MBPFORM}(b)]. The emergence of 
a nonvanishing  
local hole spin density, while the total hole spin density remains zero, is characteristic for a spin 
pseudosinglet\cite{Oszwaldowski2011:PRL,Oszwaldowski2012:PRB}, which can be understood from
a simple perturbation picture. 
The exchange interaction 
admixes higher (single particle) orbitals to the ground-state $s$ orbital. 
Specifically, as shown in the Fig.~\ref{fig:MBPFORM}(c), the mixing of $s$ and $p_x$ orbitals leads
to the spin-Wigner molecule,\cite{Oszwaldowski2012:PRB} a spin-analog of the Wigner 
molecule.\cite{Yannouleas2007:RPP,Ghosal2006:NP,Egger1999:PRL,Ellenberger2006:PRL,Singha2010:PRL,Schinner2013:PRL} 
In contrast to the Wigner molecule, where the spatial carrier separation originates from the Coulomb repulsion,\cite{Egger1999:PRL}
here the dominant contribution of such separation is typically the carrier-Mn exchange energy.

The corresponding pseudosinglet wave function\cite{Oszwaldowski2011:PRL} at each snapshot of Mn configuration, 
 $\{\xi_k \}$, is 
\begin{equation}
\phi_{\text{PS}}(r_1,r_2)=\frac{ N}{\sqrt{2}}\left[u(r_1)d(r_2)\ket{\uparrow; \downarrow}
- u(r_2)d(r_1)\ket{\downarrow; \uparrow}\right],
\label{eq:Psudosing}
\end{equation}
where $N= 1/(1+\varepsilon_x^2)$ is the normalization constant, $u(r,\{\xi_k \}) = s(r) + \varepsilon_x(\{\xi_k \}) p_x(r)$,
$d(r,\{\xi_k \}) = s(r) - \varepsilon_x(\{\xi_k \}) p_x(r)$, and $\varepsilon_x(\{\xi_k \})$ is a mixing (variational) 
parameter that depends 
on $\{\xi_k \}$.  In choosing our variational wave function we neglect overlaps between like-orbitals, 
($s-s$ and $p_x - p_x$), which results in the loss of fluctuations in the total magnetization at the site of Mn. 
These fluctuations are small in the MBP regime. 

The variational energy $E(\varepsilon_x)$ is 
\begin{equation}
E = \frac{E_s + \varepsilon_x ^2 E_p}
{1+\varepsilon_x ^2}+U - \frac{\beta}{3} \sum_{j=1}^{N_\text{Mn}} \rho_{\text{MBP}}(R_j,\varepsilon_x) S_{j,z},
\label{eq:Etotv2}
\end{equation}
where the first term is the sum of the kinetic and potential energy, the second term is the 
Coulomb energy, taken to be constant, and the third term is the average exchange energy between the MBP spin density, 
\begin{equation}
\rho_{\text{MBP}}(R_j,\varepsilon_x) = \frac{ 6\varepsilon_x }{1+\varepsilon_x^2} s(R_j) p_x(R_j),
\label{eq:MBPSPINDENS}
\end{equation}
at the site of Mn.  
For a nonmagnetic system, the two-particle energy for the ground state is  
$E_s = 2 \hbar \omega$ and for the $p$-state is $E_p = 4 \hbar \omega$.
To get close to the eigenstate of the system, 
we seek the $\varepsilon_x$ that minimizes Eq.~(\ref{eq:Etotv2})
\begin{equation}
\varepsilon_{x,\text{min}}= 
\frac{\Delta_{\rm MBP}}{E_{ps}/2+ \sqrt{E _{ps}^2/4 + \Delta_{\text{MBP}}^2}},
\label{eq:exmin}
\end{equation}
 and minimized energy, 
\begin{equation}
E_{\text{min}} = \frac{1}{2} \left(  E_s +E_p - \sqrt{E _{ps}^2 + 4 \Delta_{\text{MBP}}^2} \right)+U, 
\label{eq:Emin3v2}
\end{equation}
where $E _{ps}= E_p -E_s$ is the energy difference between the $s$ 
and $p_x$ orbital, and the spin splitting due to orbital polarization,
\begin{equation}
\Delta_{\text{MBP}} =\beta  \sum_{j=1}^{N_\text{Mn}}s(R_j) p_x(R_j)S_{j,z}=E _{ps} \sum_{k=1}^{N_C} \chi_k \xi_k ,
\label{eq:sigma}
\end{equation}
with a dimensionless magnetic susceptibility like term 
\begin{equation}
 \chi_k = \frac{S N_k \beta s(R_k) p_x(R_k)}{E _{ps}}.
\label{eq:sigmadiscrete}
\end{equation} 

To obtain the full partition function we need to sum over all configurations of $\xi_k$.  
Since we are in the $k_{\text{B}}T \ll E _{ps}$ regime, we consider only the spin 
``singlet'' ground state and neglect the possibility of a triplet state.\cite{note:triplet} 
The partition function is
\begin{equation}
Z_{\text{MBP}}=\int e^{- E_{\text{min}}/k_BT} \prod_{k=1}^{N_c}
\Omega_S(N_k, \xi_k)d^{N_c}\xi.
\label{eq:PartMBP1}
\end{equation}

For simplicity, to evaluate Eq.~(\ref{eq:PartMBP1}), we use a two site model 
by placing $N_k$ Mn at two opposite sites 
equally spaced from the origin along the $x$-axis.  
The sites are chosen such that  $s(R_1)p_x(R_1) = -s(R_2)p_x(R_2)$, 
where $R_1$ and $R_2$ are the position of the Mn ions
at site 1 and site 2, respectively. For this two site problem, Eq.~(\ref{eq:sigma}) 
reduces to $\Delta_{\text{MBP}} =E _{ps} \overline{\chi} \xi_{-}$, where
\begin{equation}
\xi_{-} =\xi_2 -\xi_1,
\label{eq:x} 
\end{equation}
and $\overline{\chi} = |\chi_1| = |\chi_2|$, $N_1=N_2=N_{\rm Mn}/2$. 

To investigate the MBP as a function of $T$, we begin by using the MFA.  
It can be shown that the minimum of the free energy, 
$F_{\text{MBP}}(\xi_{-} ) = E_{\text{min}}(\xi_{1},\xi_{2} )+ G_S(\xi_{1}) +G_S(\xi_{2})$,
lies on the $\xi_1 = -\xi_2$ line. 
The free energy 
becomes
\begin{equation}
F_{\text{MBP}}(\xi_{-} ) = E_{\text{min}}(\xi_{-} )+ 2 G_S(\xi_{-} /2).
\label{eq:FreeMBP}
\end{equation}
Minimization of the MBP free energy, 
Eq.~(\ref{eq:FreeMBP}), with respect to $\xi_{-}$
gives a self-consistent equation for  $\Delta^{\rm MF}_{\text{MBP}}$, 
\begin{equation}
\frac{ \Delta^{\rm MF}_{\text{MBP}}  }{2}=E_{ps} \overline{\chi} B_S\left(\frac{4 \overline{\chi} \Delta^{\rm MF}_{\text{MBP}}}{ N_{\rm Mn} k_{\text{B}}T \sqrt{E^2_{ps}+4  (\Delta^{\rm MF}_{\text{MBP}})^2}} \right),
\label{eq:orderMBP}
\end{equation}
where $\Delta^{\rm MF}_{\text{MBP}} =E _{ps}  \overline{\chi} \xi^{\rm MF}_{-}$.
In the unsaturated limit, $\Delta^{\rm MF}_{\text{MBP}} /k_B T \ll 1$, Eq.~(\ref{eq:orderMBP}) gives
a vanishing $\Delta^{\rm MF}_{\text{MBP}} $ at a critical temperature,
\begin{equation}
k_\text{B} T_{C,{\rm MBP}}^{\rm MF}=\frac{8S(S+1)  \overline{\chi}^2 E_{ps}}{3 N_{\rm Mn}} \propto \beta^2. 
\label{eq:TC2}
\end{equation}

Unlike the MP, Eq.~(\ref{eq:MPMFSELFCONTC}), the MF critical temperature for MBP is quadratic in $\beta$, through $\overline{\chi}^2$,  
a result similar to that of degenerate holes in bulk DMS, Eq.~(\ref{eq:TCBULKdeg}).

\begin{figure}[h]
\centerline{\psfig{file=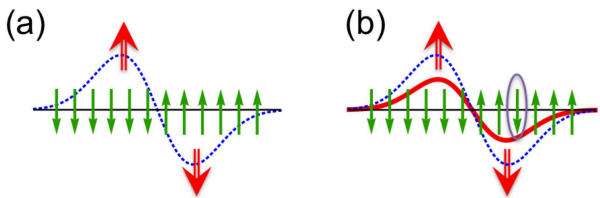,width=1.0\linewidth,angle=0}} 
\caption{(Color online)~Illustration of the finite-size effect for the MBP 
(a) Spin density (blue/dashed) for fully aligned Mn spins (green). 
(b) A flipped Mn spin changes the carriers spin density (red/solid).   }
\label{fig:figMBPFINITEEF}
\end{figure}
 We see that $\rho_{\text{MBP}}(\varepsilon_x)$ in Eq.~(\ref{eq:MBPSPINDENS}) 
 is coupled linearly to the magnetic ordering 
 of the Mn spins via Eqs.~(\ref{eq:exmin}) and ~(\ref{eq:sigma}), 
 a result that is analogous to Pauli paramagnetism where the 
 magnetization of a free electron gas is proportional to the strength of an external magnetic field.  
 Thus, any small change in the Mn configuration will result in a linear response 
 in the distribution of the carriers' spin density.  The strength of the response is determined by a 
 Pauli-like susceptibly term given by Eq.~(\ref{eq:sigmadiscrete}). Figure~\ref{fig:figMBPFINITEEF}(a)
shows that one Mn spin ``sees'' the carrier orbital spin while one carrier spin ``sees'' all Mn collectively. 
Since the height of the spin density is determined by $\varepsilon_x$, 
which is dependent on the configuration of Mn spins, if one Mn spin changes,
Fig.~\ref{fig:figMBPFINITEEF}(b), the exchange field arising from the Mn field 
is strong that the carrier responds to the change.
 Therefore, inverting one Mn spin, Fig.~\ref{fig:figMBPFINITEEF}(b), 
will change the amplitude of the spin density. Through the exchange interaction between 
the holes' spin density and Mn spins, 
all other Mn will respond to the changing spin density. 
Thus, the Mn are indirectly interacting with one another through the hole.
 
\begin{figure}[h]
\centerline{\psfig{file=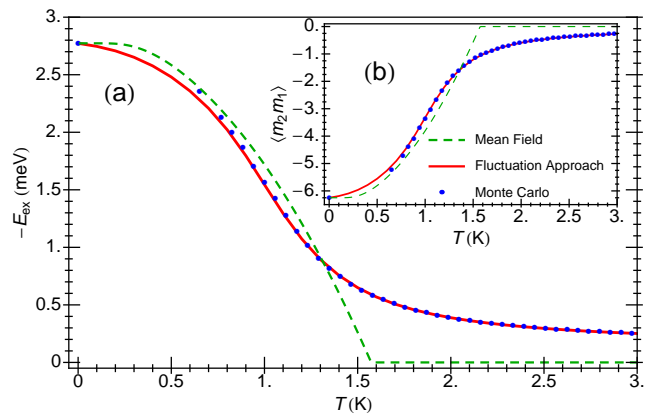,width=1.0\linewidth,angle=0}} 
\caption{(Color online)~(a) 
$T$-dependence of the average exchange energy 
and the removal of the phase 
transition for the two site model of the MBP, 
[Eq.~(\ref{eq:EexMFMBP})] with a non-fluctuating carrier spin density
(green/dashed) and a fluctuating spin density (red/solid) [Eq.~(\ref{eq:Eexfull2site})]. 
(b) 
$T$-dependence of the product of magnetization of each site, 
$\langle m_2m_1\rangle$, for a fluctuating spin 
density (red/solid) and nonfluctuating spin 
density (green/dashed). 
The parameters are $h_z=2.5 \text{ nm}$, $\hbar \omega = 30 \text{ meV}$, $m^*_h =0.21$, 
$N_0\beta$ = $-$1.05 eV, $U = 30$ meV, $R_{1,2} = \pm2$~nm and $N_{\rm Mn} = 10$.
}
\label{fig:figMBPFAMF}
\end{figure}

The average exchange energy for the MBP 
is obtained the same way as the MP,
using the solutions of Eq.~(\ref{eq:orderMBP}) to give
\begin{equation}
E_{\text{ex}}^{\text{MF}} = \frac{- (\Delta^{\rm MF}_{\text{MBP}})^2 }{  \sqrt{E^2_{ps}/4 +(\Delta^{\rm MF}_{\text{MBP}})^2}}.
\label{eq:EexMFMBP}
\end{equation}
The green/dashed line in Fig.~\ref{fig:figMBPFAMF}(a) shows the 
$T$-dependence of Eq.~(\ref{eq:EexMFMBP}),
while Fig.~\ref{fig:figMBPFAMF}(b) shows the mean field behavior of the averaged product 
of the normalized magnetization at the two sites, $\langle m_2 m_1 \rangle$. 
There is an antiferromagnetic correlation between the product of normalized magnetization at the two sites.  
As a consequence of thermal spin fluctuations, the average exchange energy and the magnitude of $\langle m_2 m_1 \rangle$ 
decreases  and with increasing  $T$ 
eventually vanishing when the MFA 
for the specific parameter set 
yields a vanishing carrier spin density resulting in a second order phase transition at $T_{C,{\rm MBP}}^{\rm MF}=1.5$~K.

Exact integration of the MBP partition function Eq.~(\ref{eq:PartMBP1}),
which correctly includes spin fluctuations,  
is needed to remove the phase transitions predicted by MF theory.
The average exchange energy is obtained through
 \begin{eqnarray}
E_{\text{ex}} &=&  \frac{1}{Z_{\text{MBP}}}  \int d\xi_1\int d\xi_2 \Omega_S(N_1, \xi_1) \Omega_S(N_2, \xi_2)  \nonumber \\ 
&\times& 
\left[\beta\frac{\partial E_{\rm
min}(\xi_1,\xi_2)}{\partial\beta}\right]
e^{- E_{\text{min}} (\xi_1, \xi_2) /k_\text{B} T}, 
\label{eq:Eexfull2site}
\end{eqnarray}
A $T$-dependence of this average exchange energy 
is shown by the red/solid lines 
in 
Fig.~\ref{fig:figMBPFAMF}(a).
With the inclusion of spin fluctuations, the phase transition is removed resulting in a finite $E_{\text{ex}}$ at  $T>T_C$.
The product of the normalized magnetization of each site
was calculated numerically to demonstrate the antiferromagnetic correlation between the Mn spins 
at the two sites. 

Monte Carlo simulations 
(see appendix~\ref{MC} for details) verify the theoretical prediction for the MBP.
$s-p$ levels coupled via Mn spins are directly diagonalized and the variational assumption, 
Eq.~(\ref{eq:Psudosing}), was not used.   
Mn ions (5 per site) were placed 2 nm from the center of the QD along the $x$-axis. 
We include $s$, $p_x$ and $p_y$  orbitals in the simulation.  The Coulomb energy, as before,
 is a constant. The blue dots in Fig.~\ref{fig:figMBPFAMF}(a) shows the $T$-dependence of the
 average exchange energy and the product of the magnetization per site is shown in Fig.~\ref{fig:figMBPFAMF}(b).
 There is an excellent
 agreement between the fluctuation approach and Monte Carlo simulations. 
 Furthermore, the antiferromagnetic correlation  between the two sites demonstrates the formation of the MBP.  
  
 As we did for the MP, we investigate two limiting cases for $E_{\text{ex}}$.
 In the saturated regime, the Mn are maximally aligned at each site with their 
 spin pointing in opposite directions with $|\xi_1-\xi_2|\to 2$. In
this limit, the MF expression, Eq.~(\ref{eq:EexMFMBP}), is valid. In the
opposite unsaturated high-$T$ limit, $e^{-E_{\rm
min}(\xi_1,\xi_2)/k_BT}\approx 1$ can be taken in
Eq.~(\ref{eq:Eexfull2site}). Finally, we obtain the limiting behavior of
$E_{\rm ex}$ within the regime $\overline{\chi}\ll 1$, 
\begin{equation}
E_{\rm ex}\approx\left\{
\begin{array}{ll}
-6\frac{S}{S+1} N_{\rm Mn}T_{C,{\rm MBP}}^{\rm MF}\propto (N_{\rm Mn}\beta)^2 & \mbox{ for }T\to 0 \\
\\
-T_{C,{\rm MBP}}^{\rm MF}\propto N_{\rm Mn}\beta^2 & \mbox{ for }T\to \infty
\end{array}
\right.
\end{equation}
This shows that, unlike the MP, the coupling dependence is always in
$\beta^2$. This is due to the fact that the carrier spin density is
polarized from zero by the coupling to Mn spins, 
unlike the uncompensated
carrier spin in the MP.
  
\begin{figure}[h]
\centerline{\psfig{file=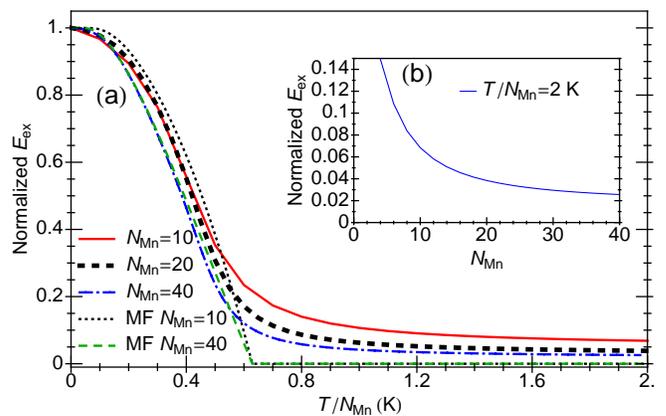,width=1.0\linewidth,angle=0}} 
\caption{(Color online) The finite-size effect in a doubly occupied QD.  
(a) $T$-dependence of the normalized MBP average exchange energy 
for 10 Mn (red/solid), 20 Mn (black/dashed) and 40 Mn (blue/dash-dotted), thus  
5, 10, and 20 Mn per site, respectively. 
Mean-field solution with 10 Mn (black/dotted) and 40 Mn (green/dashed).  
(b) The normalized 
mean-field solution with varying number of Mn 
for a fixed $T/N_{\rm Mn}$.   
}
\label{fig:EMBPNMN}
\end{figure}

The MBP displays a finite-size effect different from the MP.  
We plot the $T$-dependence of the 
average exchange energy, Eq.~(\ref{eq:Eexfull2site}), 
normalized by its low-$T$ value in Fig.~\ref{fig:EMBPNMN}(a) with a varying number of Mn in a given cell. 
As the number of Mn increases, the normalized $E_{\text{ex}}$ fluctuation tail decays toward the MF solution.  
This is demonstrated in Fig.~\ref{fig:EMBPNMN}(b) at a fixed ratio of $T/N_{\rm Mn}$. As the number of Mn ion spins increases
toward the thermodynamic limit, the normalized exchange energy decays. 
This size dependence is expected from the MF result where a phase 
transition to zero exchange energy occurs. The $1/N_{\rm Mn}$ dependence in 
$E_{\rm ex}$ reflects the thermodynamic limit towards the vanishing order parameter.

To summarize, like in the MP case, the 
MFA can be used to understand the
statistical properties of the MBP. Applying the variational treatment to the carrier
spins, we express the exchange energy as $E_{\text{ex}} = -a(\xi_1 -\xi_2)^2$ 
with the Mn spins $\xi_{1,2}$ at position 1 and 2.  
This quadratic dependence $(\xi_1 - \xi_2)^2$  
 originates from the carrier spin being linearly polarized out of the spin-singlet  
in closed-shell systems.
With the variable $X \equiv \xi_1 -\xi_2$, the free energy 
becomes
\begin{equation}
F_{\text{MBP}}(\xi_1,\xi_2) \approx -a X^2 - T S(\xi_1,\xi_2).
\label{eq:FreeMBPeq1}
\end{equation}

Confining the discussions to unpolarized states $(\xi_1+ \xi_2 = 0)$, the entropy becomes
an even function of $X$ and Eq.~(\ref{eq:freenergyfunc}) results with $g_1 =0$ [see
Fig.~\ref{fig:figINTRO1}(b)]. 
We note that, in contrast to the Eq.~\ref{eq:freenergyfunc},
here the first $X^2$ term in the free energy is not of an entropic origin.
In a finite system $\langle X \rangle = 0$ and the exchange energy
$E_{\text{ex}}= - a \langle X^2 \rangle$  
 remains finite. While the MFA 
 incorrectly predicts
a phase transition to $E_{\text{ex}}=0$ beyond a phase transition temperature, it correctly
justifies the finite-size scaling limit $ E_{\text{ex}}/N_{\rm Mn} \rightarrow 0$ as $N_{\rm Mn} \rightarrow \infty$
as shown in
Fig.~\ref{fig:figINTRO1}(d), in a sharp contrast to the MP case. 
The open or closed-shell electronic
structures 
leads 
to fundamentally different statistical properties in the quantum
dot magnetism.

 \section{Conclusions}
  
 Many of our findings for magnetic polarons and bipolarons can be also applied to higher carrier occupancy of quantum dots
 where it is important to distinguish if they form open- or closed-shell systems which leads to a qualitatively different classes 
 of magnetic ordering. Performing a thermodynamic analysis in these nanoscale magnets, 
 we reveal the limitations 
 of a mean-field approximation and the necessity for a more accurate theoretical framework that would correctly 
 include spin fluctuations. 
 Our results show that 
 a careful choice of the order parameter in the mean-field approximation (using the exchange 
 energy, rather than magnetization) removes spurious phase transitions for magnetic polarons, but not for magnetic 
 bipolarons. In the later case,  the phase transitions  are removed by including spin fluctuations within the 
 coarse-grained method and Monte Carlo simulations. 

The conventional mean-field theory, known from bulk systems with nondegenerate or 
degenerate carrier density, reveals important differences between the magnetic polarons and bipolarons. 
Surprisingly, we can introduce a very simple mean-field form of a free energy to accurately describe qualitatively 
different finite-size effects and distinct thermodynamic limits in magnetic polarons and bipolarons with the change 
of the number of magnetic impurity spins.  
These findings remain unchanged once we carefully include 
spin fluctuations, further justifying our simple description and a pictorial difference between the finite-size effects in 
 magnetic polarons and bipolarons, Figs.~\ref{fig:figMPFINITEEF} and \ref{fig:figMBPFINITEEF}, respectively. 
 
Similar to our prediction for an unexpected thermally-enhanced magnetic ordering in quantum dots,\cite{Pientka2012:PRB}
a judicious use of the mean-field approximation and awareness of its artifacts could provide important insights in unexplored 
 aspects of nanoscale magnets.
 For example, we expect that the mean-field description of the different finite-size 
 scaling in magnetic order discussed for magnetic polaron and bipolaron will also apply to other open- and closed 
 shell quantum dots with higher carrier occupancy. 
 A mean-field calculation of the critical temperature 
 could also reveal a different power-law dependence in the exchange coupling constant for the exchange energy of 
 open- and closed-shell systems. 
 
In contrast to magnetic polarons, 
 much less is known about 
 magnetism in closed-shell systems, often simply 
 implying that the magnetic ordering is completely absent. Therefore, to test our predictions for magnetic 
 bipolarons it would be important to focus on the experimental realization of multiple carrier occupancy in quantum dots. 
 The simple creation of excitons  is not sufficient. A simultaneous presence of single electron and hole 
 effectively just renormalizes the exchange coupling with magnetic impurities 
 of magnetic polarons.\cite{Yakovlev:2010} 
 Instead, photoexcitiation, 
 using chemical and electrostatic doping,\cite{note_field} 
 should create a pair of holes or electrons. 
 
 Another possibility would be to fabricate quantum dots from novel Mn-doped II-II-V dilute magnetic semiconductors. 
 These systems provide an independent charge and spin doping and would therefore be suitable to test formation of 
 nanoscale magnetism for a wide range of parameters.\cite{Zhao2013:NC,Zhao2014:CSB,Glasbrenner2014:PRB} 
 They share with (II,Mn)VIs an isovalent character of Mn-doping, removing the solubility 
 constraint of (III,Mn)Vs an obstacle for 
 fabricating magnetic quantum dots.\cite{Kudelsk2007:PRL} 
 Unlike (II,Mn)VI, Mn-doped II-II-V systems can separately attain different carrier densities through independent 
 charge doping and thus readily alter the strength of the Mn-carrier exchange coupling.

\section{acknowledgements} 
We 
thank Peter Stano for  discussions about
the stability of magnetic bipolarons and Alex Matos Abiague for his
valuable comments. 
This work was primarily supported by the U.S. DOE-BES,
Office of Science,  under Award DE-SC0004890 (I.\v{Z}.), as well as grants 
U.S. ONR N000141310754 (J.P.), and NSF-DMR 0907150 (J.H).
\appendix
\section{ Mn Distribution Function} \label{MnDist}
We discretize 
the QD space with $N_k$ magnetic moments 
in a given cell.  
The distribution function is found by summing over all configurations of the Mn in a 
given cell and is determined by Eq.~(\ref{eq:OMEGA}). 
Using the 
integral representation of the 
$\delta$-function, $\delta(x) = (1/2 \pi) \int _{- \infty}^{\infty} d\lambda e^{-i \lambda x}$, 
the partition function is 
 \begin{equation}
\Omega_S(N_k,\xi_k)=\frac{N_k S}{2 \pi} \sum_{\left\{  S_{jz}\right\}  }\ \int _{- \infty}^{\infty} d\lambda e^{-  i \lambda  \left( N_k S \xi_k -\sum_{j=1}^{N_k} S_{jz} \right) }
\label{eq:Ymn2}
\end{equation}  
Introducing a complex variable $h = i \lambda S$, 
Eq.~(\ref{eq:Ymn2}) 
becomes
 \begin{equation}
\Omega_S(N_k,\xi_k) =\frac{N_k }{2 \pi i }\int _{-i \infty}^{i\infty} e^{ N_k \left( \ln Z_S(h)    - h\xi_k \right) } dh, 
\label{eq:Ymn4}
\end{equation} 
where $Z_S(x) = \sinh\left[ (1+1/2S) x \right] /\sinh\left[ x /2S \right]$.  
The integrand in Eq.~(\ref{eq:Ymn4} )
is sharply peaked (Gaussian-like), therefore we can approximate Eq.~(\ref{eq:Ymn4}) 
by performing the method of steepest descent.
We deform the contour in the complex plane to pass through a saddle point in the 
direction of steepest descent.  
By Taylor expansion of the function in the exponent of Eq.~(\ref{eq:Ymn4}) and performing a Gaussian integral
over $h$ we obtain 
\begin{equation}
\Omega_S(N_k,\xi_k) = \sqrt{\frac{N_k}{2 \pi \chi (h_k)} }e^{ -G_S (\xi_k,T) / k_B T },
\label{eq:Ymn6}
\end{equation} 
where $\chi (h) =  \partial^2 \ln Z_S(h)/\partial h^2$ and $G_S (\xi_k,T)$ 
is the Gibbs free energy, recall Eq.~(\ref{eq:MnFree}),
obtained through a Legendre transformation.\cite{Petukhov2007:PRL}

\section{Magnetic Polaron Renormalization} \label{RENORM} 
Since the MP is not influenced by the finite-size effect (recall Figs.~\ref{fig:figMPFINITEEF} and \ref{fig:figEMPNMN}),
we can employ a variational approach to study the
influence of the wave function renormalization on the MP properties in a QD. 
Starting from the MP partition function
in  Eq.~(\ref{eq:PartMP2}), we approximate the MP wave function by a single $s$-orbital of a 2D harmonic oscillator,
constant over the the QD height, $h_z$,
\begin{equation}
\phi(r) =1/(\sqrt{h_z \pi} L_{\text{MP}})e^{-(x^2 +y^2)/2L_{\text{MP}}^2}.
\label{eq:renormWF}
\end{equation}
Using a variational approach\cite{Dietl1982:PRL,Wolff:1988} we
determine the width, $L_{\text{MP}}$, 
that minimizes the 
total free energy functional, $F_{\text{MP}}$.  
The MP 
free energy functional is given by
\begin{eqnarray}
F_{\text{MP}}&=& \frac{\hbar \omega}{2} \left(\frac{L_0^2}{L_{\rm MP}^2} + \frac{L_{\rm MP}^2}{L_0^2} \right)   -k_{\text{B}} T \ln 2 \nonumber\\ 
&-&k_{\text{B}} T \sum_{j=1}^{N_\text{Mn}}\ln \left[ Z_S\left(\frac{\beta S \rho_{\mathrm{MP}}(R_j)}{ 3 k_{\text{B}} T } \right)   \right],
 \label{eq:FreeMP1}
\end{eqnarray} 
where the first two terms are the sum of the kinetic and potential energy
 with $L_0 = \sqrt{\hbar/m^* \omega}$, 
the third term is from the hole spin degeneracy, and the final term is due
to the exchange interaction of the hole spin density $\rho_{\mathrm{MP}} (r)$ 
at the site of Mn spins, $R_j$. As an approximation, we consider a homogeneous distribution of Mn,
and transform  $\sum_j \rightarrow N_0 x_{\text{Mn}} \int d^3R$
(the continuous limit)
where $x_{\rm Mn}$ is the Mn fraction per cation and $N_0$ is the density of cation sites. 
The average exchange interaction for the MP, $E_{\text{MP}}$,
can be derived to obtain
 \begin{equation}
E_{\text{MP}} = - \frac{\Delta_{\text{max}}}{3} \int d^3R  \rho_{\mathrm{MP}}(R)B_{S}\left(\frac{ \beta S \rho_{\text{MP}}(R)}{3 k_{\text{B}} T}   \right), 
\label{eq:EMPint}
\end{equation}
where $\Delta_{\text{max}} =x_{\text{Mn}} |N_0 \beta| S $.
 $E_{\text{MP}}$ is found by numerically minimizing Eq.~(\ref{eq:FreeMP1})
to obtain the most probable width, $L_{\text{MP}}$. 
This width determines $\rho_{\text{MP}}$ by combining Eqs.~(\ref{eq:renormWF}) and (\ref{eq:rhoMP}) and 
yields   $E_{\text{MP}}$ from Eq.~(\ref{eq:EMPint}).

\begin{figure}[h]
\centerline{\psfig{file=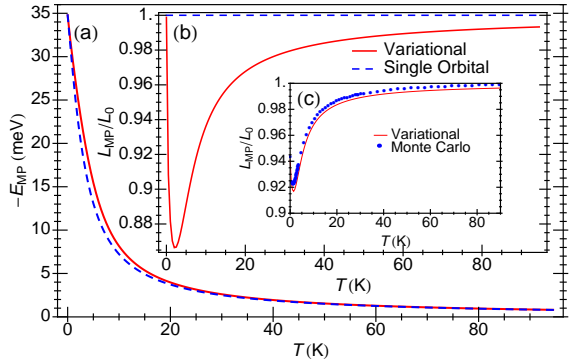,width=0.9\linewidth,angle=0}} 
\caption{(Color online)
(a)  
$T$-dependence of the MP energy [Eq.~(\ref{eq:EMPint})]  with the
variationally-obtained   (``renormalized'') (red/solid) and the fixed wave function for a
harmonic QD confinement without Mn spins. 
(b) The corresponding 
$T$-dependence of the most probable width for a renormalized wave 
function,  $L_{\text{MP}}$ (red/solid) and fixed wave function (blue/dashed), $L_{\text{MP}}\equiv L_0$, from Eq.~(\ref{eq:renormWF}). 
(c) 
$T$-dependence of the $L_{\text{MP}}$ 
from a variational method (red/solid) and Monte Carlo simulations (blue/dotted).
}
\label{fig:figRenorm}
\end{figure}

It is instructive to now compare how various forms of the carrier wave function affect the $T$-dependence of the
 $E_{\text{MP}}$, shown in Fig.~{\ref{fig:figRenorm}}(a). We choose 
$x_{\text{Mn}}=2.6 \%$, $h_z=2.5 \text{ nm}$,
$\hbar \omega = 30 \text{ meV}$, $m^*_h =0.21$, $L_0 = 3.5 \text{ nm}$--the characteristic width in the absence of Mn spins, 
and $N_0\beta$ = $-$1.05 eV, which within the classical radius of this harmonic confinement yields $N_{\text{Mn}}=90$,
as in Fig.~\ref{fig:figMFFAMP}. 
From Fig.~{\ref{fig:figRenorm}}(a), where we compare our results for the variationally obtained $L_{\text{MP}}(T)$ with the wave function of a fixed width at $L_{\text{MP}}(T)\equiv L_0$, 
we see that the 
wave function renormalization have a very small influence on $E_{\text{MP}}(T)$. 
In fact, both of them are very similar to the variational $E_{\text{MP}}(T)$ for a constant wave function in Fig.~\ref{fig:figMFFAMP}.

From Fig.~{\ref{fig:figRenorm}}(b) we see that the 
wave function
renormalization itself is a small effect. 
The red/solid curve shows how the wave function width, normalized to the nonmagnetic width,
$L_0$, varies with $T$. 
As the $T$ increases from $T=0$~K, the Mn spins coupled to the tail of the wave function are 
more prone to thermal excitation. The wave function  shrinks to attain a 
more energetically favorable configuration by increasing the exchange energy gain from
the polarized Mn spins near the center of the QD.  Eventually, thermal excitation 
overcomes the magnetic energy, and the system relaxes continuously to a 
nonmagnetic state resulting in $L_{\text{MP}}=L_0$ at large $T$, 
as there is no energy gain from the wave function renormalization. 
From the variational approach we see an additional MP localization: $L_{\text{MP}}\leq L_0$, while the 
nonmonotonic $L_{\text{MP}}(T)$ implies also a nonmonotonic effective exchange field\cite{Yakovlev:2010}
due to the MP formation.

To further verify the renormalization effect of the MP wave function, we implement Monte Carlo 
simulations, removing the need of a variational calculations of the wave function.  
To approximate the MP wave function we include  $s$, $p$, and $d$ orbitals in our simulation 
and allow for mixing of these orbitals.  
In Fig.~\ref{fig:figRenorm}(c) we see that both  variational (red/solid) and Monte Carlo (blue/dotted) results  
agree well with each other and shown again a nonmonotonic  $L_\text{MP}(T)$, noted also in  Fig.~\ref{fig:figRenorm}(b). 
The slightly smaller wave function renormalization in Fig.~\ref{fig:figRenorm}(c), as compared to that  in 
Fig.~\ref{fig:figRenorm}(b), is a consequence of fewer Mn spins in the middle region of the QD.

\section{Monte Carlo Simulations} \label{MC} 
Monte Carlo simulations  were 
used to approximate solutions to the Schr\"{o}dinger equation 
\begin{equation}
\hat{H}(\{S\}) \ket{\Phi} = E(\{S\})  \ket{\Phi},
\end{equation}
for a fixed finite orthonormal basis $\ket{\Phi}$ at a given Mn spin $(\{S\})$ configuration. 
The calculation entails guessing a Mn configuration at a given $T$,  
producing a matrix
representation of $\hat{H}(\{S\})$ in a finite basis, and solving 
the eigenvalue problem.

The calculation begins by  defining a 2D harmonic 
QD and solve for single heavy hole 
QD levels and eigenfunctions without any Mn atoms.  We truncate the states up to the 
first $N$-orbitals. We obtain $N$ noninteracting wave functions $\phi_{n \sigma}^0({\bf r})$ at energy $E_n^0$ 
with $\sigma = \uparrow, \downarrow$.
For a given configuration of $N_{\text{Mn}}$ Mn spins, \{$S_{z1},S_{z2},...,S_{z N_{\text{Mn}}}$\}, 
we construct a $(2N) \times (2N)$ matrix,
\begin{equation}
\hat{H} = \sum_{n \sigma} E_n^0 \ket{\phi_{n \sigma}^0} \bra{\phi_{n \sigma}^0}+ \sum_{nn', \sigma} 
g_{nn'\sigma}
(\{ S_z\}) \ket{\phi_{n' \sigma}^0} \bra{\phi_{n \sigma}^0}. 
\end{equation}
 The interaction constant $g$ is
\begin{equation}
g_{nn'\sigma}
(\{ S_z\}) = \frac{\beta}{3} \sum_{j=1}^{N_{\text{Mn}}} [\phi_{n' \sigma}^0({\bf R}_j)]^* 
\left(s_{z,\sigma}
S_{zj}\right) \phi_{n \sigma}^0({\bf R}_j), 
\end{equation}
where $s_z$ is the spin of the carrier and ${\bf R}_j$ is the position of the Mn ion.
We diagonalize the single heavy hole Hamiltonian at a snapshot of $\{ S_z\}$
and obtain 
eigenvalues $E_i(\{ S_z\})$ and eigenvectors $c_{i,n\sigma}(\{ S_z\})$.
We then propose a different Mn-configuration, \{$S'_{z1},S'_{z2},...,S'_{z N_{\text{Mn}}}$\},
diagonalize $\hat{H} ( \{ S'_z\})$ and obtain new eigenvalues $E'_i(\{ S'_z\})$.


\begin{thebibliography}{10}

\bibitem{Chudnovski1988:PRL}
E.~M. Chudnovsky and L.~Gunther, 
Phys. Rev. Lett. {\bf 60}, 661 (1988).

\bibitem{Friedman1996:PRL}
J.~R Friedman, M. P.~Sarachik, J.~Tejada, and R.~Ziolo,
Phys. Rev. Lett. {\bf 76}, 3830 (1996).

\bibitem{Garanin1997:PRB}
D.~A. Garanin and E.~M. Chudnovsky,
Phys. Rev. B {\bf 56}, 11102 (1997). 

\bibitem{Seufert2001:PRL}
J.~Seufert, G.~Bacher, M.~Scheibner, A.~Forchel, S.~Lee, M.~Dobrowolska, and J.~K. Furdyna,
Phys. Rev. Lett. {\bf 88}, 027402 (2001). 

\bibitem{Beaulac2009:S}
R.  Beaulac, L. Schneider, P.~I. Archer, G. Bacher, and D.~R. Gamelin,
Science {\bf 325}, 973  (2009). 

\bibitem{Fernandez-Rossier2007:PRL} 
J.~Fern\'andez-Rossier and R.~Aguado,
Phys. Rev. Lett. {\bf 98}, 106805 (2007).

\bibitem{Abolfath2008:PRL}
R.~M. Abolfath, A.~G. Petukhov, and I. \v{Z}uti\'c, 
Phys. Rev. Lett. {\bf 101}, 207202 (2008).

\bibitem{Abolfath2007:PRL}
 R.~M. Abolfath,  P. Hawrylak,  and I. \v{Z}uti\'c, 
Phys. Rev. Lett. {\bf 98}, 207203 (2007).

\bibitem{Oszwaldowski2012:PRB}
R.~Oszwaldowski, P.~Stano, A.G.~Petukhov, and I. \v{Z}uti\'c, 
Phys. Rev. B {\bf 86}, 201408(R) (2012).

\bibitem{Wasner:2012}
R. Wasner (Ed.) {\em Nanoelectronics and Information Technology} 3$^{\text{rd}}$ Edition (Wiley,  Hoboken,  2012).

 \bibitem{Raman2013:N}
K.~V. Raman, A.~M. Kamerbeek, A. Mukherjee, N. Atodiresei, T.~K. Sen, P. Lazi\'c, V. Caciuc, R. Michel, D. Stalke, S. K. Mandal, 
S. Mandal, S. Blugel, M. Munzenberg, J. S. Moodera     
Nature {\bf 493}, 509 (2013).

\bibitem{Chen2014:NN}
J.-Y. Chen, T.-M. Wong, C.-W. Chang, C.-Y. Dong, and Y.-F. Chen,
Nat. Nanotech. {\bf 9}, 845 (2014).
 
\bibitem{Zutic2014:NN}
I. \v{Z}uti\'c and P. E. Faria Junior, 
Nat. Nanotech. {\bf 9}, 750 (2014).

\bibitem{DiVincenzo 1995:S}
D.~D. DiVincenzo, Science {\bf 279}, 255 (1995).

\bibitem{Thiele2014:S}
S. Thiele, F. Balestro, S. Klyatskaya, M. Ruben, W. Wernsdorfer,
Science {\bf 344}, 1135 (2014).

\bibitem{Maksimov2000:PRB}
A.~A. Maksimov, G.~Bacher, A.~McDonald, V.~D. Kulakovskii, A.~Forchel, C.~R. Becker, 
G.~Landwehr, and L.~W. Molenkamp,
Phys. Rev. B  {\bf 62}, R7767 (2000).

\bibitem{Dorozhkin2003:PRB}
P.~S. Dorozhkin,  A.~V. Chernenko, V.~D. Kulakovskii, A.~S. Brichkin, A.~A. Maksimov, 
H. Schoemig, G. Bacher, A. Forchel, S. Lee, M. Dobrowolska, and J.~K. Furdyna,
Phys. Rev. B  {\bf 68}, 195313 (2003).

\bibitem{Holub2004:APL}
M. Holub, S. Chakrabarti, S. Fathpourand,  P. Bhattacharya, Y. Lei, and S. Ghosh,
Appl. Phys. Lett.  {\bf 85}, 973 (2004).

\bibitem{Besombes2004:PRL}
L. Besombes, Y. Leger, L. Maingault, D. Ferrand, H. Mariette, and J. Cibert, 
Phys. Rev. Lett.  {\bf 93}, 207403 (2004).

\bibitem{Hundt2005:PRB}
A. Hundt, J. Puls, A. V. Akimov, Y. H. Fan, and F. Henneberger, Phys. Rev. B 72, 033304 (2005).

\bibitem{Gurung2008:APL}
T. Gurung, S. Mackowski, G. Karczewski, H.~E. Jackson, and L.~M. Smith,
Appl. Phys. Lett. {\bf 93}, 153114 (2008).

\bibitem{Hoffman2000:SSC}
D.~M. Hoffman, B.~K. Meyer, A.~I. Ekimov, I.~A. Merkulov, Al.~L. Efros, M. Rosen, G. Couino, 
T. Gacoin, and J.~P. Boilot, Solid State Commun. {\bf 114}, 547 (2000).

\bibitem{Norris2001:NL}
D. J. Norris, N. Yao,  F.~T. Charnock, and T. ~A. Kennedy, Nano Lett. {\bf 1}, 3 (2001).

\bibitem{Radovanovic2001:JACS} 
P.~V. Radovanovic and D.~R. Gamelin, J. Am. Chem. Soc.  {\bf 123}, 12207( 2001).

\bibitem{Ochsenbein2009:NN}
S.~T. Ochsenbein, Y. Feng,    K.~M. Whitaker, E. Badaeva,  W.~K  Liu, X. Li,   and D.~R. Gamelin,                              
Nat. Nanotech. {\bf 4},  68 (2009).

\bibitem{Zutic2009:NN}
I. \v Zuti\'c and A. G. Petukhov,
Nat. Nanotech.  {\bf 4}, 623 (2009).

\bibitem{Bussian2009:NM}
D.~A. Bussian, S.~A. Crooker, M. Yin, M. Brynda, A.~L. Efros, and V.~I. Klimov, 
Nat. Mater. {\bf 8}, 35 (2009).

\bibitem{Viswanatha2011:PRL}
R. Viswanatha, J.~M. Pietryga, V.~I. Klimov, and S.~A. Crooker,
Phys. Rev. Lett. {\bf 107}, 067402 (2011).

\bibitem{Beaulac2011:PRB}
R. Beaulac, Y. Feng, J.~W. May, E. Badaeva, D.~R. Gamelin, X. Li., 
Phys. Rev. B {\bf 84}, 195324 (2011).

\bibitem{Pandey2012:NN}
A. Pandey, S. Brovelli, R. Viswanatha, L. Li, J.~M. Pietryga, V.~I. Klimov, and S.~A. Crooker, 
Nat. Nanotech. {\bf 7} , 792 (2012).


\bibitem{Farvid2014:JACS}
S.~S. Farvid, T. Sabergharesou, L.~N. Huffluss, M. Hegde, E. Prouzet, and P.~V. Radovanovic, 
J. Am. Chem. Soc. {\bf 136}, 7669 (2014).

\bibitem{Peng2014:JPC}
B. Peng, J.~W. May, D.~R. Gamelin, and X. Li, 
J. Phys. Chem. {\bf 118}, 7630 (2014).

\bibitem{Xiu2010:ACSN}
F. Xiu, Y. Wang, J. Kim, P. Upadhyaya, Yi Zhou, X. Kou, W. Han, R.~K. Kawakami, J. Zou, and K.~L. Wang.
ACS Nano  {\bf 4}, 4948  (2010).

\bibitem{Sellers2010:PRB}
I.~R. Sellers, R. Oszwa\l{}dowski, V.~R. Whiteside, M. Eginligil, A.~Petrou, I. \v{Z}uti\'{c}, 
W.-C. Chou, W.~C. Fan, A.~G. Petukhov, S.~J. Kim, A.~N. Cartwright, and B.~D. McCombe,
Phys. Rev. B  {\bf 82}, 195320 (2010);

\bibitem{Barman2015:PRB}
B. Barman,  R. Oszwa\l{}dowski, L. Schweidenback, A.~H. Russ, J.~M. Pientka, Y. Tsai,  W-C. Chou, W.~C. Fan,
J.~R. Murphy, A.~N. Cartwright, I.~R. Sellers, A.~G. Petukhov, I. \v{Z}uti\'{c},  B.~D. McCombe, and A. Petrou,
Phys. Rev. {\bf 92}, 035430 (2015).

\bibitem{Henneberger:2010}
F. Henneberger and J. Puls, in {\it Introduction to the
Physics of Diluted Magnetic Semiconductors} edited by
J. Kossut and J.~A. Gaj (Springer, Berlin, 2010).

\bibitem{Klopotowski2011:PRB}
 \L{}. K\l{}opotowski,  \L{}. Cywi\'{n}ski,  P. Wojnar, V. Voliotis, K. Fronc, T. Kazimierczuk, 
A. Golnik, M. Ravaro, R. Grousson, G. Karczewski, and T. Wojtowicz,
Phys. Rev. B {\bf 83}, 081306(R) (2011).

\bibitem{Klopotowski2013:PRB}
 \L{}. K\l{}opotowski,  \L{}. Cywi\'{n}ski, M. Szymura, V. Voliotis, R. Grousson, P. Wojnar, K. Fronc, T. Kazimierczuk, 
 A. Golnik, G. Karczewski, and T. Wojtowicz,
Phys. Rev.  B {\bf 87}, 245316 (2013).

\bibitem{Pacuski2014:CGD}
W. Pacuski,T. Jakubczyk, C. Kruse, J. Kobak, T. Kazimierczuk, M. Goryca, A. Golnik,
P. Kossacki, M. Wiater, P. Wojnar, G. Karczewski, T. Wojtowicz, and D. Hommel,
Cryst. Growth Des. {\bf 14}, 998 (2014). 

\bibitem{Kudelsk2007:PRL} %I32 moved from the end
A. Kudelski, A. Lema"tre, A. Miard, P. Voisin, T. ~C.~M. Graham, R.~J. Warburton, and O. Krebs, 
Phys. Rev. Lett. {\bf 99}, 247209 (2007)

\bibitem{vanBree2008:PRB}%I32 added
J. van Bree, P.~M. Koenraad, and J. Fernandez-Rossier, 
Phys. Rev. B {\bf 78}, 165414 (2008).

\bibitem{Govorov2004:PRB} %I32 added
A.~O. Govorov, Phys. Rev. B {\bf 70}, 035321 (2004).

\bibitem{Yakovlev:2010}
D.~R. Yakovlev and W.Ossau,  in
{\em Introduction to the Physics of Diluted Magnetic Semiconductors}
edited by J. Kossut and J.~A. Gaj (Springer, Berlin, 2010).

\bibitem{Dietl1982:PRL}
T.~Dietl and J.~Spa\l{}ek, Phys. Rev. Lett.  {\bf 48}, 355 (1982);
T.~Dietl and J.~Spa\l{}ek, Phys. Rev. B  {\bf 28}, 1548  (1983).

\bibitem{Wolff:1988} 
P.~A. Wolff,  in  {\em Semiconductors and Semimetals} edited by J.~K. Furdyna and
J. Kossut  (Academic Press, San Diego 1988), Vol. 25.

\bibitem{Durst2002:PRB}
A.~ C. Durst,  R.~N. Bhatt, and P.~A. Wolff,
Phys. Rev. B  {\bf 65}, 235205 (2002).

\bibitem{Furdyna1988:JAP} 
J.~K. Furdyna,  J. Appl. Phys.  {\bf 64} R29 (1988).

\bibitem{Nagaev:1983}   
E.~L. Nagaev,  {\em Physics of Magnetic Semiconductors}
(MIR Publishers, Moscow 1983).

\bibitem{Kasuya1968:RMP}
T. Kasuya and A. Yanase,
Rev. Mod. Phys. {\bf 40}, 684 (1968).

\bibitem{Dietl1995:PRL} 
T. Dietl, P. Peyla, W. Grieshaber, and Y. Merle d' Aubign\'{e},
Phys. Rev. Lett. {\bf 74}, 474 (1995).

\bibitem{Awschalom1986:PRL}
D.~D. Awschalom, J. Warnock, and S. von Moln\'{a}r,
Phys. Rev. Lett. {\bf 58}, 812 (1987).


\bibitem{Marder:2010}
M.~P. Marder {\em Condensed Matter Physics} 2$^{\text{nd}}$ Edition (Wiley, Hoboken, 2010). 

\bibitem{Free_note}
Clearly nanomagnets are finite and thus spatially-inhomogeneous systems which require
using free energy functional or a closely related effective Hamiltonian [Ref.~\onlinecite{Dietl1982:PRL}],
not  a spatially-independent MFA free energy, $F$. However, we %X will 
explain 
that a 
careful choice of the order parameter in $F$ can yield important trends in nanomagnets. 

\bibitem{temp_note}
For a carrier-mediated magnetism in QDs %X quantum dots 
this description would break down above 
$T$ for which the thermal energy exceeds the confinement energy leading to no carriers. 

\bibitem{Govorov2005:PRB}
A.~O. Govorov,
Phys. Rev. B.  {\bf 72}, 075359 (2005).

\bibitem{Govorov2008:CRP}
A.~O. Govorov,
C.~R. Physique {\bf 9}, 857 (2008).

\bibitem{Cheng2008:PRB} %I32
S.-J. Cheng, Phys. Rev. B.  {\bf 77}, 115310 (2008).

\bibitem{Oszwaldowski2011:PRL}
R. Oszwa\l{}dowski, I. \v{Z}uti\'{c}, and A.~G. Petukhov,
Phys. Rev. Lett.  {\bf 106}, 177201 (2011).

\bibitem{Bednarski2012:JPCM}
H. Bednarski and J.~Spa\l{}ek, 
J. Phys.: Condens. Matter {\bf 24}, 235801 (2012).
 
\bibitem{Abolfath2012:PRL}
R. ~M. Abolfath, M. Korkusinski, T. Brabec, and P. Hawrylak
Phys. Rev. Lett. {\bf 108}, 247203 (2012).

\bibitem{Kuskovsky2007:PRB}
 I.~L. Kuskovsky, W. MacDonald, A.~O. Govorov, L. Mourokh, X. Wei, M.~C. Tamargo, M. Tadic, and F.~M. Peeters,
Phys. Rev. B {\bf 76}, 035342 (2007).

\bibitem{Cherenko2010:PSSB}
A.~V. Chernenko, A.~S. Brichkin, S.~V. Sokolov, and S.~V. Ivanov,
Phys. Status Solidi B {\bf 247}, 1514 (2010).

\bibitem{Gould2006:PRL} %I32
C. Gould, A. Slobodskyy, D. Supp, T. Slobodskyy, P. Grabs, P. Hawrylak, F. Qu, G. Schmidt, and L.~W. Molenkamp, 
Phys. Rev. Lett. {\bf 97}, 017202 (2006).

\bibitem{Matsuda2007:APL} 
K.~Matsuda, S.~V. Nair, H.~E. Ruda, Y.~Sugimoto, T.~Saiki, and K.~Yamaguchi,
Appl. Phys. Lett. {\bf 90}, 013101 (2007).

\bibitem{Bansal2009:PRB}
B. Bansal, S.~Godefroo, M.~Hayne, G.~Medeiros-Ribeiro, and V.~V. Moshchalkov,
Phys. Rev. B {\bf 80}, 205317 (2009).

\bibitem{Fisher2005:PRL}
B. Fisher, J.~M. Caruge, D. Zehnder, and M.~Bawendi, 
Phys. Rev. Lett.  {\bf 94}, 087403 (2005).

\bibitem{Besombes2005:PRB}
L.~Besombes, Y.~Leger, L.~Maingault, D.~Ferrand, H.~ Mariette, and 
J.~Cibert,
Phys. Rev. B {\bf 71}, 161307(R) (2005).

\bibitem{Trojnar2013:PRB}
A.~ H. Trojnar, M.  Korkusinski, U.~ C. Mendes, M. Goryca, M. Koperski, T. Smolenski, P. Kossacki, P. Wojnar, and P. Hawrylak,
Phys. Rev. B {\bf 87}, 205311 (2013).

\bibitem{DasSarma2003:PRB}
S. Das Sarma, E. H. Hwang, and A. Kaminski
Phys. Rev. B {\bf 67}, 155201 (2003).

\bibitem{Petukhov2007:PRL}
A. G. Petukhov, I.~\v{Z}uti\'{c}, and S.~C. Erwin,
Phys. Rev. Lett. {\bf 99}, 257202 (2007).

\bibitem{Pientka2012:PRB}
J. M. Pientka, R. Oszwa\l{}dowski, A. G. Petukhov, J. E. Han and I. \v{Z}uti\'{c}
Phys. Rev. B {\bf 86}, 161403(R) (2012).

\bibitem{Fernandez-Rossier2004:PRL} 
J.~Fern\'andez-Rossier and L.~Brey,
Phys. Rev. Lett. {\bf 93}, 117201 (2004).

\bibitem{Lebedeva2010:PRB}
N. Lebedeva, A. Varpula, S. Novikov, and P. Kuivalainen,
Phys. Rev. B {\bf 81}, 235307 (2010).
 
 \bibitem{Lebedeva2012:PSSB}
N. Lebedeva, A. Varpula, S. Novikov, and P. Kuivalainen,
Phys. Status Solidi B {\bf 249}, 2244 (2012).

\bibitem{Binder:2009}
D.P.~Landau, K.~Binder {\em A Guide to Monte Carlo Simulations in Statistical Physics 3rd Edition } 
(Cambridge University Press,Cambridge, 2009).

\bibitem{Lee2014:PRB}
J. Lee, K. Vyborny, J. E. Han, and I. \v{Z}uti\'c, 
Phys. Rev. B {\bf 89}, 045315 (2014)

\bibitem{Stano2013:PRB}
P. Stano, J. Fabian, and I. \v{Z}uti\'c, 
Phys. Rev. B {\bf 87}, 165303 (2013).

\bibitem{Vyborny2012:PRB}
K. Vyborny, J. E. Han, R. Oszwa\l{}dowski, I.~\v{Z}uti\'{c}, and A. G. Petukhov,
Phys. Rev. B {\bf 85}, 155312 (2012).

\bibitem{Gall2011:PRL}
C. Le Gall, A. Brunetti, H. Boukari, and L. Besombes
Phys. Rev. Lett. {\bf 107}, 057401 (2011).

\bibitem{Klosowsk1991:JAP}
P. K\l{}osowski, T. M. Giebu\l{}towicz, J. J. Rhyne,
N. Samarth, H. Luo, and J. K. Furdyna, J. Appl. Phys.
{\bf 70}, 6221 (1991). 

\bibitem{note_open} 
By defnition, the open-shell systems have more of either ``spin-up" or ``spin-down" carriers.

\bibitem{Kubo:1960}
R.~Kubo, {\em Statistical Mechanics} 
(North-Holland, Amsterdam, 1960).

\bibitem{Golnik1983:JPC}
A. Golnik, J.~Ginter, and J.~A. Gaj,
J. Phys. C
{\bf 16}, 6073 (1983). 

\bibitem{Yannouleas2007:RPP}
C. Yannouleas and U. Landman, 
Rep. Prog. Phys. {\bf  70}, 2067 (2007).

\bibitem{Ghosal2006:NP}
A. Ghosal, A.~D. Guclu, C.~J. Umrigar, D. Ullmo, and H.~U. Baranger, 
Nat. Phys. {\bf 2}, 336 (2006). 

\bibitem{Egger1999:PRL}
R. Egger, W. H\"{a}usler, C.~H. Mak, and H. Grabert, 
Phys. Rev. Lett.  {\bf 82}, 3320 (1999).

\bibitem{Ellenberger2006:PRL}
C. Ellenberger, T. Ihn, C. Yannouleas, U. Landman, K. Ensslin, D. Driscoll, and A. C. Gossard, 
Phys. Rev. Lett. {\bf 96}, 126806 (2006).

\bibitem{Singha2010:PRL}
A. Singha, V. Pellegrini, A. Pinczuk, L.~N. Pfeiffer, K. W. West, and M. Rontani, 
Phys. Rev. Lett. {\bf 104}, 246802 (2010).

\bibitem{Schinner2013:PRL}
G.~J. Schinner, J. Repp, E. Schubert, A.~K. Rai, D. Reuter, A.~D. Wieck,
A. O. Govorov, A.~W. Holleitner, and J.~P. Kotthaus,
Phys. Rev. Lett. {\bf 110}, 127403 (2013). 


\bibitem{note:triplet}
This is a very accurate assumption for relatively small Mn content we
consider here. The triplet state becomes relevant for a higher Mn content.\cite{Oszwaldowski2012:PRB} 

\bibitem{note_field}
Even electric-field-induced redistribution of carriers could yield striking
changes in magnetic ordering. For example, Ref.~\onlinecite{Abolfath2008:PRL}; E. Dias Cabral, M. A. Boselli,
R. Oszwa{\l}dowski, I. \v Zuti\'c, and I. C. da Cunha Lima, Phys. Rev. B {\bf 84},
085315 (2011); L.~D. Anh, P.~N. Hai, Y. Kasahara, Y. Iwasa, and M. Tanaka,
arXiv:1503.02174, preprint.

\bibitem{Zhao2013:NC}
K. Zhao, Z. Deng, X. C. Wang, W. Han, J. L. Zhu, X. Li, Q. Q. Liu, R. C. Yu, T. Goko, 
B. Frandsen, L. Liu, F. Ning, Y. J. Uemura, H. Dabkowska, G. M. Luke, H. Luetkens, 
E. Morenzoni, S. R. Dunsiger, A. Senyshyn, P. B\"{o}ni, and C. Q. Jin, 
Nat. Commun. {\bf 4}, 1442 (2013).

\bibitem{Zhao2014:CSB}
K. Zhao, B. Chen, G. Zhao, Z. Yuan, Q. Liu, Z. Deng, J. Zhu, and C. Jin, 
Chin. Sci. Bull. {\bf 59}, 2524 (2014).

\bibitem{Glasbrenner2014:PRB} 
J. K. Glasbrenner, I. \v{Z}uti\'c, and I. I. Mazin, 
Phys. Rev. B {\bf 90}, 140403(R) (2014).

\end{thebibliography}
\end{document}